\documentclass[11pt,a4paper]{article}
\pdfoutput=1
\usepackage{jheppub}
\usepackage{rotating}
\usepackage{array}
\usepackage{amsmath}
\usepackage[normalem]{ulem}
\usepackage{slashed}
\usepackage{booktabs}
\usepackage[pdftex,table]{xcolor}
\usepackage{units}
\usepackage{booktabs}
\usepackage[final]{pdfpages}

\newcommand{\ds}{{\sf DarkSUSY }}
\newcommand{\GENIE}{{\sf GENIE }}
\newcommand{\mZp}{m_{Z'}}
\newcommand{\gZp}{g_{Z'}}
\newcommand{\mN}{m_N} 
\newcommand{\mA}{m_A} 
\newcommand{\A}{A} 
\newcommand{\mchi}{m_\chi}
\newcommand{\Nthr}{N_\chi^{\text{thr}}}
\newcommand{\sigSI}{\sigma_p^{\text{SI}}}
\newcommand{\sigSD}{\sigma_p^{\text{SD}}}
\renewcommand{\sec}{section~}
\newcommand{\Refe}{ref.~}
\newcommand{\Refes}{refs.~}
\newcommand{\eq}{eq.~}
\newcommand{\eqs}{eqs.~}
\newcommand{\fig}{figure~}
\newcommand{\deff}{D_{\text{eff}}}

\newcommand{\cm}{\mathrm{c.m.}}

\usepackage{multirow}

\title{Prospects of detecting cosmic ray up-scattered dark matter with DUNE
}

\author[1]{Richard Diurba}
\author[2]{$\!\!$, and Helena Kole\v{s}ov\'{a}}

\affiliation[1]{Albert Einstein Center, Laboratory for High Energy Physics, University of Bern, Sidlerstrasse 5, CH-3012 Bern, Switzerland}
\affiliation[2]{Department of Mathematics and Physics, University of Stavanger, 4036 Stavanger, Norway}

\emailAdd{richard.diurba@unibe.ch}
\emailAdd{helena.kolesova@uis.no}

\abstract{%

Detection of sub-GeV dark matter (DM) particles in direct detection experiments is inherently difficult, as their low kinetic energies in the galactic halo are insufficient to produce observable recoils of the heavy nuclei in the detectors. On the other hand, whenever DM particles interact with nucleons, they can be accelerated by scattering with galactic cosmic rays. These cosmic-ray-boosted DM particles can then interact not only through coherent elastic scattering with nuclei, but also through scattering with individual nucleons in the detectors and produce outgoing particles at MeV to GeV kinetic energies. The resulting signal spectrum overlaps with the detection capabilities of modern neutrino experiments. One future experiment is the Deep Underground Neutrino Experiment (DUNE) at the Sanford Underground Research Facility. Our study shows that DUNE has a unique ability to search for cosmic-ray boosted DM with sensitivity comparable to dedicated direct detection experiments in the case of spin-independent interactions. Importantly, DUNE's sensitivity reaches similar values of DM-nucleon cross sections also in the case of spin-dependent interactions, offering a key advantage over traditional direct detection experiments.

}

\arxivnumber{2504.16996}


\begin{document}
\maketitle
\flushbottom

\section{Introduction}

The nature of dark matter (DM) in our Universe remains one of the most profound mysteries in modern physics. New particles beyond the Standard Model (SM) are often proposed to explain the existence of DM, and various strategies have been developed to probe their nature~\cite{Bertone:2004pz}. 
In this work, we focus on the so-called direct detection of dark matter, with rare interactions between DM particles and atomic nuclei in terrestrial detectors observed as the Earth moves through the DM halo surrounding our galaxy~\cite{Goodman:1984dc}. 
Most attention in direct detection experiments goes towards the example of spin-independent (SI) interactions with nuclei, which occurs, e.g., through DM coupling to SM quarks via a scalar or vector mediator. The SI interactions of DM with nucleons within a nucleus are coherently summed, hence, the DM-nucleus cross section scales with the square of the atomic number $A^2$. Consequently, experiments utilizing large-$A$ xenon nuclei, such as LZ~\cite{LZ:2022lsv} or XENONnT~\cite{XENON:2023cxc}, have achieved remarkable sensitivity to DM-nucleon cross sections as low as $\sigSI \sim 10^{-47}\,$cm$^2$ for DM masses around $30\,$GeV. However, kinematic constraints limit the sensitivity of these experiments to low-mass DM, as particles from the galactic halo have velocities constrained by the galactic escape velocity of $\sim 540\,$km/s and lack the kinetic energy needed to produce a detectable recoil on a heavy nucleus. As a result, extensive efforts are underway to enhance experimental sensitivity to hadrophylic halo dark matter in the sub-GeV mass range. New detection strategies with lower detection threshold are being developed~\cite{Essig:2022dfa}, or alternative processes related to DM-nucleus scattering with more favorable kinematics are being considered, such as the production of Migdal electrons~\cite{Ibe:2017yqa,XENON:2019zpr,DarkSide:2022dhx} or bremsstrahlung photons~\cite{Kouvaris:2016afs}. 

Another strategy is to explore scenarios where DM is accelerated by specific mechanisms, allowing sub-GeV DM to acquire enough kinetic energy to produce a detectable nuclear recoil in existing direct detection experiments or 
to produce signatures in neutrino experiments. A flux of accelerated DM coming to Earth may arise, for example, when a heavier dark sector component annihilates or decays into a lighter one~\cite{Agashe:2014yua, Berger:2014sqa, Kopp:2015bfa}, when DM undergoes scattering with particles in the Sun \cite{Kouvaris:2015nsa, An:2017ojc, Emken:2017hnp}, or with energetic galactic cosmic-ray (CR) nuclei~\cite{Yin:2018yjn,Bringmann:2018cvk,Cappiello:2018hsu}. In this work, we concentrate on the last option of cosmic-ray up-scattered dark matter (CRDM). Although acceleration by CR electrons and subsequent detection via interaction with electrons in different detectors was considered in some studies~\cite{Ema:2018bih,Dent:2020syp,Dent:2020qev,Bardhan:2022ywd,Cao:2020bwd,Jho:2020sku,PandaX:2024pme,Herbermann:2024kcy,Guha:2024mjr}, we focus on the case of hadrophylic DM studied in the context of CRDM by \Refes\cite{Dent:2019krz,Wang:2019jtk,Bondarenko:2019vrb,Cho:2020mnc,Guo:2020oum,Xia:2022tid,Cappiello:2019qsw,Ema:2020ulo,Ge:2020yuf,Harnik:2020ugb,Bramante:2021dyx,PandaX-II:2021kai,Chauhan:2021fzu,Super-Kamiokande:2022ncz,Alvey:2022pad,Kolesova:2022kvq,Maity:2022exk,Nagao:2022azp,NEWSdm:2023qyb,Wang:2023wrx,Bell:2023sdq,Su:2023zgr,Dutta:2024kuj,Reis:2024wfy,Diurba:2024dqo,LZ:2025iaw}.  
While CRDM constraints from direct detection experiments are based on the coherent elastic scattering of DM with nuclei, it was shown in~\cite{Cappiello:2019qsw,Ema:2020ulo,Chauhan:2021fzu,Super-Kamiokande:2022ncz} that neutrino experiments may detect the scattering of DM with individual nucleons, referred to as quasi-elastic (QE) scattering in this text. Furthermore, scattering with individual partons, so-called deep inelastic scattering (DIS), was shown to be crucial for possible CRDM detection in IceCube~\cite{Guo:2020drq,Guo:2020oum,Cappiello:2024acu}.\footnote{Interestingly, \Refes\cite{Dutta:2024kuj,Lin:2025pqh} suggest further processes that may lead to detection of accelerated DM in neutrino detectors, however, these are not considered in our work.}

From the experiments considered in \Refes\cite{Ema:2020ulo,Super-Kamiokande:2022ncz,Guo:2020oum,Cappiello:2024acu}, the Deep Underground Neutrino Experiment (DUNE) has the highest sensitivity to DM with mass in the MeV-GeV range that is most interesting for us, as lower DM masses are typically strongly constrained by the Big Bang Nucleosynthesis (BBN)~\cite{Krnjaic:2019dzc}. With four planned modules with 17.5 kilotons of liquid argon~\cite{DUNE:2020lwj}, DUNE can provide the exposure and detector capabilities to search for DM boosted by different mechanisms~\cite{Kim:2016zjx,Berger:2019ttc,DeRoeck:2020ntj,Acevedo:2024wmx}, including CRDM. This work focuses on this experiment and improves upon existing works in several aspects. First, we employ more precise modeling of the interactions relevant for DM detection. In particular, \Refe\cite{Ema:2020ulo} modeled the QE process as scattering off free nucleons and neglected the DIS contribution when determining the sensitivity region of the DUNE detector. Our work employs the boosted dark matter module~\cite{Berger:2018urf} of the \GENIE interaction event generator~\cite{Andreopoulos2009rq,Andreopoulos:2015wxa,GENIE2021wox} to properly model the complicated nuclear effects that influence the QE scattering and add DIS. 
Second, this paper studies the effect of the atmospheric neutrino background on the DUNE sensitivity, which was neglected in~\cite{Ema:2020ulo}. Finally, the inelastic scattering is newly considered for the DM up-scattering by CR. The effect of DIS in CR-DM scattering was discussed in~\cite{Guo:2020oum}; however, they focused on the secondary photons produced in such a process. The enhancement of DM up-scattering due to inelastic scattering was discussed in~\cite{Su:2023zgr,Diurba:2024dqo} focusing on the implications for direct detection experiments, where the bounds are not enormously extended since these experiments predominantly detect DM with lower kinetic energies. However, the sensitivity of neutrino experiments can be significantly enhanced by inclusion of the inelastic CR-DM scattering, as we show in the present work. 

The original work~\cite{Bringmann:2018cvk} attempted to present the CRDM constraints in a model independent way, assuming the simplified ``constant'' cross section for DM scattering with nucleons, thereby, neglecting the dependence on the transferred momentum $Q^2$. On the other hand, follow-up works like~\cite{Dent:2019krz,Wang:2019jtk,Bondarenko:2019vrb,Cho:2020mnc,Ema:2020ulo,Super-Kamiokande:2022ncz,Alvey:2022pad,Wang:2023wrx,Bell:2023sdq,Su:2023zgr,Dutta:2024kuj,Diurba:2024dqo} showed that the $Q^2$ dependence might alter the CRDM constraints for different DM scenarios. Moreover, in order to describe the DIS process, a specific model for DM-quark interactions must be chosen. This work is based on DM-quark interactions mediated by a new massive gauge boson $Z'$ as implemented in the \GENIE event generator~\cite{Berger:2018urf}. We consider the vector couplings that lead to SI scattering with nuclei mentioned above and axial-vector couplings that lead to spin-dependent (SD) interactions. For SD scattering, the interactions with DM are stronger for nuclei with convenient spin structures, like the fluorine isotope $^{19}$F, and there is no $A^2$ enhancement for nuclei with large atomic numbers. Therefore, the direct detection bounds on the DM-nucleon cross section based on halo dark matter are relatively weaker for SD scattering than for SI scattering, at only $ \sigSD \sim 10^{-41}\,$cm$^2$ for $\mathcal{O}(10\,$GeV) DM masses~\cite{PICO:2019vsc}. 
In contrast, we show that the CRDM constraints for SI and SD cases remain relatively similar. In fact, CRDM constraints are expected to exist for any model where DM couples to nucleons since both the boosting and detection mechanisms rely solely on the DM-nucleon interaction.

This work will first introduce the mechanism of DM up-scattering by interactions with CR (\sec\ref{sec:CRflux}) and discuss the detection of CRDM particles (\sec\ref{sec:detection}). Example DM scenarios are described in \sec\ref{sec:dmint}, where the resulting interactions with nucleons and nuclei are also discussed. The corresponding sensitivities for DUNE are then presented and compared with complementary constraints in \sec\ref{sec:results}. We conclude in \sec\ref{sec:conc} and discuss several technical aspects in the appendices.

\section{Cosmic-ray up-scattered dark matter} \label{sec:crdm}
In this section, we will explain the mechanism of DM acceleration by galactic CR and the calculation of the possible event rate in the DUNE experiment. The description of the attenuation of the CRDM flux is postponed to appendix~\ref{sec:att}.
\subsection{Cosmic-ray up-scattering}\label{sec:CRflux}
Let us consider galactic CR nuclei $\A$ with kinetic energies $T_\A$ and differential fluxes $d\Phi_\A/dT_\A(x)$ that vary by position. These energetic particles may scatter with halo DM with density $\rho_\chi(x)$, giving rise to a flux of up-scattered DM particles with kinetic energies of $T_\chi$~\cite{Bringmann:2018cvk}:\footnote{The CRDM flux~\eqref{eq:CRflux1} is defined as the total flux coming to Earth from all directions, whereas other works like~\cite{Xia:2021vbz} give the CRDM flux per unit solid angle and their result, hence, differ from ours by a factor of $4\pi$. The factor of $1/(4\pi)$ appears in~\eqref{eq:CRflux1} since $d\Phi_\A/dT_\A$ is the CR flux integrated over all directions; therefore, division by $4\pi$ is needed to obtain the averaged CR flux per unit solid angle.}
\begin{eqnarray} \label{eq:CRflux1}
\frac{d\Phi_{\chi}}{dT_\chi}&=&\int\frac{d\Omega}{4\pi} \int_{\rm l.o.s.} \!\!\!\!\!\!d\ell \, \frac{\rho_\chi}{m_\chi} 
\sum_\A
\int_{T_\A^\mathrm{min}}^\infty d T_\A\, \frac{d \sigma_{\chi \A} }{dT_\chi} \frac{{d\Phi_\A}}{dT_\A}\\
&\equiv&  
\deff \frac{\rho_\chi^\mathrm{local}}{m_\chi}  
\sum_\A
\int_{T_\A^\mathrm{min}}^\infty d T_\A\, \frac{d \sigma_{\chi \A} }{dT_\chi} \frac{{d\Phi^\mathrm{LIS}_\A}}{dT_\A}
\label{eq:CRflux2}
\,.
\end{eqnarray}
On the second line, the dependencies of the DM density and CR flux on position are captured by the effective distance $\deff$. Therefore, only the local interstellar CR flux, $d\Phi^\mathrm{LIS}_\A/dT_\A$ \cite{Boschini:2018baj,Boschini:2020jty} and the local DM density, $\rho_\chi^\mathrm{local} = 0.3\,$GeV/cm$^3$, appear in the final formula. We assume $\deff = 10\,$kpc in this work, as motivated by the detailed analysis presented in \Refe\cite{Xia:2021vbz}.

 Furthermore, $d \sigma_{\chi \A}/dT_\chi$ in \eq\eqref{eq:CRflux1} is the differential cross section for the scattering of CR nuclei with DM at rest where $T_\chi$ is the kinetic energy of the recoiled DM particle. We improve existing results by adding the effect of inelastic CR-DM scattering that may play an important role in scenarios where elastic scattering is suppressed. 
 As mentioned above, we extract the cross sections for inelastic DM-nucleus scattering from the numerical code \GENIE~\cite{Berger:2018urf}, and these are calculated in the nucleus rest frame. 
 Conveniently, the Lorentz-invariant momentum transfer $Q^2$ is related to $T_\chi$ by
\begin{equation}\label{eq:defQCR}
Q^2 = 2 m_\chi T_\chi,
\end{equation} leading to $d \sigma_{\chi \A}/dT_\chi = 2 \mchi d \sigma_{\chi \A}/dQ^2$. The latter cross section is invariant under the boost between the DM and nucleus rest frames, allowing the \GENIE results to be directly used. Let us note that two kinematic variables are needed to describe inelastic scattering, hence, $d \sigma_{\chi \A}/dQ^2$ in fact refers to an integral, such as:
\begin{equation}\label{eq:doublediff}
\frac{d \sigma_{\chi \A}}{dQ^2} \equiv \int_{X} \frac{d^2 \sigma_{\chi \A}}{dQ^2 dX} dX
\end{equation}
where $X$ refers to some other kinematic variable (see examples in appendix~\ref{ap:kinematics}). When inferring $d \sigma_{\chi \A}/dQ^2$ from~\GENIE, the DM kinetic energy in the nucleus rest frame $\tilde{T}_{\chi,\mathrm{in}}$ is needed. The following relation between $\tilde{T}_{\chi,\mathrm{in}}$ and the kinetic energy of the CR nucleus $T_\A$ in the DM rest frame is then used:
\begin{equation}\label{eq:tildeTchi}
\tilde{T}_{\chi,\mathrm{in}} = \frac{m_\chi}{m_\A} T_\A.
\end{equation}
Finally, the lower bound for the integration over the CR kinetic energies in \eq\eqref{eq:CRflux1}, $T_A^{\min}$, arises because a given final DM kinetic energy $T_\chi$ can be obtained only for sufficiently large initial CR kinetic energy. Different processes considered for CR-DM scattering involve different kinematics (see appendix~\ref{ap:kinematics}). However, it turns out that $T_A^{\min}$ attains the smallest value for the case of coherent DM-nucleus scattering, where
\begin{equation}\label{eq:TAmin}
    T_A^{\min} = \frac{T_\chi}{2}-\mA + \sqrt{\left(\frac{T_\chi}{2}-\mA\right)^2 + \frac{(\mchi + \mA)^2 T_\chi}{2\mchi}}.
\end{equation}

In practice, we evaluate the CRDM flux using the 6.4 version of the \ds code~\cite{Gondolo:2004sc,Bringmann:2018lay,Bringmann:2022vra}. We benefit from the fact that user modification can be easily implemented and we complement the public version of \ds by including inelastic CR-DM scattering.\footnote{Let us note that inelastic scattering of DM with nuclei is implemented in the public version of \ds for the purposes of CRDM flux attenuation. This implementation is based on~\cite{Alvey:2022pad} where the output of the \texttt{GiBUU} code~\cite{Buss:2011mx,gibuuweb} for neutrino-nucleus inelastic cross sections is rescaled in order to estimate the corresponding DM-nucleus cross sections. In our work we use the inelastic cross sections calculated directly for our DM models using the \GENIE code~\cite{Berger:2018urf}, both for CR-DM scattering and for CRDM flux attenuation.}

The CRDM flux~\eqref{eq:CRflux2} is expected to reach the top of the Earth's atmosphere. Still, DM with relatively strong interactions may have part of the flux reduced by interactions with the Earth's atmosphere or crust. To model such flux attenuation in a fully reliable way, dedicated Monte Carlo simulations are needed~\cite{Emken:2017qmp} and performing these is beyond the scope of this work. We, therefore, restrict our signal region to DM particles that come ``from above'', more precisely, from zenith angles $\theta_z$ satisfying $\cos\theta_z \geq 0.1$. We checked that when this cut is applied, the number of events in DUNE should not be affected by the CRDM flux attenuation for the smallest cross sections that DUNE is capable to probe. Consequently, the main results of our paper are unaffected by the stopping of CRDM particles. We still studied this effect in a simplified way to obtain a rough estimate of the upper bound of the DUNE sensitivity region, as described in appendix~\ref{sec:att}.


\subsection{Detection methodology}\label{sec:detection}

The CRDM flux coming to the Earth's atmosphere~\eqref{eq:CRflux2} is assumed to be roughly isotropic in this work.\footnote{The anisotropy of the CRDM flux related to the larger DM density in the galactic center~\cite{Ge:2020yuf, Super-Kamiokande:2022ncz, Nagao:2022azp, NEWSdm:2023qyb} may help with background subtraction as we also comment on in \sec\ref{sec:conc}.} The attenuation of this flux depends on the thickness of the soil that must be penetrated to reach the detector, that is, on the direction of the incoming DM particle. The CRDM flux at the DUNE location $\Phi_{\text{\tiny{DUNE}}}$ is, therefore, anisotropic and when calculating the possible DM event rate in DUNE, integration over the zenith angle is performed:
\begin{equation}\label{eq:rateDUNE}
R_\chi = N_{\text{Ar}} \int_{T_\chi}  dT_\chi \, \sigma_{\chi \text{Ar}}^{\text{eff}}(T_\chi) \int_{0.1}^1 d\cos \theta_z \frac{d\Phi_{\text{\tiny{DUNE}}}}{dT_\chi d\cos\theta_z}  \,.
\end{equation}
Here $N_{\text{Ar}} = 40\,$kiloton$/m_{\text{Ar}}$ is the number of argon nuclei in the DUNE detector, with $m_{\text{Ar}} = 39.95 \times $AMU (atomic mass unit). Further, $\sigma_{\text{Ar}\chi}^{\text{eff}}$ is the effective total cross section for scattering of a DM particle with kinetic energy $T_\chi$ off an argon nucleus, taking into account the DUNE detection thresholds from~\cite{DUNE:2024qgl} (see appendix~\ref{secGENIE} for details regarding the detection thresholds).

The most important background relevant to the scattering of energetic DM particles in DUNE is from neutral-current interactions of atmospheric neutrinos. We estimate the number of background atmospheric neutrino events in appendix~\ref{sec:atmnu}. We arrive at a threshold number of DM events at $\Nthr = 196.7$, needed to observe DM at the $2\sigma$ level. Only statistical errors are considered for the number of background neutrino events that enter into our estimate of $\Nthr$.

\section{Dark matter interactions} \label{sec:dmint}
\subsection{Model for dark matter-quark interaction}
A specific model has to be chosen already at the quark level to describe energetic DM interactions. Utilizing the \GENIE module for boosted DM \cite{Berger:2018urf}, we present an example scenario with a massive $U(1)_D$ gauge boson $Z'$ assumed. It couples to a DM fermion $\chi$ via
\begin{equation}
\mathcal{L}_{\chi,\text{int}}=g_{Z'} Z_\mu' \overline{\chi} \gamma^\mu \left(Q^L_\chi P_L + Q^R_\chi P_R\right)\chi.
\end{equation}
The new mediator also interacts with SM fermions as:
\begin{equation}
\mathcal{L}_{f,\text{int}}=g_{Z'} Z_\mu' \overline{\psi}_f \gamma^\mu \left(Q^L_f P_L + Q^R_f P_R\right)\psi_f
\end{equation}
where, in practice, only the four lightest quarks $f = u,\,d,\,s,\,c$ contribute in our analysis. In the above expressions, $P_L$ and $P_R$ are projectors on the left- and right-handed components of a fermion, respectively: $P_{L/R} = (1\mp \gamma^5)/2$, and $g_{Z'}$ is the gauge coupling.

For the simplest case of the vector mediator, we choose that
\begin{equation}\label{eq:vector_charges}
Q^L_\chi = Q^R_\chi \equiv Q_\chi^V \,,\qquad   Q^L_q = Q^R_q \equiv Q_q^V \, \qquad\text{(vector couplings)}.
\end{equation}
If the $Q_q^V$ charges are equal to the electric charges, this case can be mapped onto the standard ``dark photon'' scenario. The $U(1)$ gauge boson that couples to SM fields via kinetic mixing with the SM photon in this scenario is relatively tightly constrained (see \Refe\cite{Fabbrichesi:2020wbt} for a review). Additionally, the charge assignment~\eqref{eq:vector_charges} also corresponds to the case where the SM baryon number symmetry $U(1)_B$ is gauged, which is much less constrained (depending on the branching fraction to visible and dark sectors, as shown through phenomenological studies in \Refes\cite{Ilten:2018crw,FileviezPerez:2020mta,Batell:2021snh,Blanco:2019hah}). In our numerical results, we choose all the charges $Q_q^V$ to be equal, corresponding rather to the gauged $U(1)_B$; however, our results are independent of the $Z'$ couplings to $b$ and $t$ quarks and also the impact of the coupling to $c$ is small since it contributes to DIS process only.

The second scenario under consideration is described as 
\begin{equation}\label{eq:axial_charges}
- Q^L_\chi = Q^R_\chi \equiv Q_\chi^A ,\qquad   - Q^L_q = Q^R_q \equiv Q_q^A  \, \qquad\text{(axial-vector couplings)}.
\end{equation}
Also in the axial-vector case, we assume the magnitude of the axial-vector charges to be equal for all relevant quarks. 

For a fully consistent model, conditions on anomaly cancellation should be satisfied, the mechanism of $Z'$ mass generation should be specified, or the issue of the Higgs charge assignment should be addressed in the axial-vector case. In this work, we choose the above models mostly for illustration. We do not elaborate further on the above-mentioned issues, we refer to~\cite{Carone:1994aa, Carone:1995pu,Aranda:1998fr, Alves:2013tqa,Tulin:2014tya,Lebedev:2014bba,Kahn:2016vjr,Dror:2017ehi,Blanco:2019hah,Frank:2020byg,Greljo:2022dwn,Alanne:2022eem,Balan:2024cmq} for more details on $Z'$ phenomenology.


\subsection{Scattering with nucleons}

In this section, we present formulas for the elastic DM-nucleon cross sections that are used directly in our work to describe DM scattering with CR protons and also serve as the basis for describing QE scattering of DM with nuclei.

Nucleon matrix elements for the two relevant quark operators can be parametrized by a set of form factors in the following way~\cite{Bishara:2017pfq}:
\begin{align}
    \langle N(p')| \bar{q}\gamma^\mu q | N(p)\rangle &= \overline{u}(p')\left(F_1^{q|N} (Q^2) \gamma^\mu + F_2^{q|N}(Q^2) \frac{i\, q_\nu \sigma^{\mu\nu}}{2m_N}\right)u(p)\, ,\\
\langle N(p')| \bar{q}\gamma^\mu \gamma^5 q | N(p)\rangle &= \overline{u}(p')\left(F_A^{q|N}(Q^2)\gamma^\mu \gamma^5 + F_P^{q|N}(Q^2)\frac{q^\mu}{2m_N}\gamma^5\right)u(p)\, .
\end{align}
Here $\sigma^{\mu\nu} = i/2 \left[\gamma^\mu,\gamma^\nu\right]$, $p$, and $p'$ are the initial and final 4-momenta of the nucleon $N$ ($N=n,p$), respectively, and $m_N$ is the nucleon mass. The term $q = p'-p$ is the 4-momentum exchanged in the collision and we denote $Q^2 \equiv -q^2$. 

With the charge assignments from \eqs\eqref{eq:vector_charges} and~\eqref{eq:axial_charges} considered in this work, the formula for the DM-nucleon cross section presented in~\cite{Berger:2018urf} simplifies to:\footnote{Minor typos were corrected when compared to the text of~\cite{Berger:2018urf}. Our formulas correspond to the output of the current version of the~\GENIE module for the DM-proton elastic cross section.}
\begin{equation}\label{dsig_dq2}
\frac{d\sigma_{\chi N}}{dQ^2}=\frac{g_{Z'}^4 Q_\chi^2 \mN^4}{\pi\, s \,Q^2_{\max} } \, \frac{1}{(Q^2+\mZp^2)^2} \left[ A + C\frac{(s-u)^2}{16\,\mN^4} \right]\,,
\end{equation}
where $s$ and $u$ are the usual Mandelstam variables. In the reference frame where the nucleon is initially at rest, $s$ is given by:
\begin{equation}\label{eq:s_Nrest}
s = (\mN + \mchi)^2 + 2 \mN T_\chi
\end{equation}
with $T_\chi$ being the initial DM kinetic energy and 
\begin{equation}
s-u = 4\mN (m_\chi + T_\chi) - Q^2\,.
\end{equation}
The maximum transferred 4-momentum is given by
\begin{equation}\label{Q2max_Nrest}
    Q^2_{\max} = \frac{4\mN^2}{s}\left(T_\chi^2 + 2\mchi T_\chi\right)\,. 
\end{equation}
The other coefficients in \eq\eqref{dsig_dq2} differ for the vector and axial-vector cases.

\paragraph{Vector couplings.}
For the charge assignment from Eq.~\eqref{eq:vector_charges}, $Q_\chi \equiv Q_\chi^V$ in \eq\eqref{dsig_dq2} and 
\begin{align}
A&=A_{11} (F_1^N)^2 + A_{12} F_1^N F_2^N + A_{22} (F_2^N)^2 \, , \\
A_{11} &= \tau \left(\tau  - \delta-1\right)\, ,\\
A_{12}&=2 \tau \left(2\tau  - \delta\right)\, ,\\
A_{22}&=\tau \left[\tau\left(1-\tau\right)   - \delta\right]\, ,\\
C&=(F_1^N)^2 + \tau (F_2^N)^2\, ,
\end{align}
with $\tau \equiv Q^2/(4\mN^2)$ and $\delta \equiv \mchi^2/\mN^2$.
The form factors are related to quark vector charges in the following way:
\begin{equation}
    F_i^{N} = \sum_{q=u,d,s} Q_q^V F_i^{q|N}\, ,
\end{equation}
$i=1,2$. Aligned with definitions in Ref.~\cite{Berger:2018urf}, this work employs the isospin limit ($F_i^{u|p} = F_i^{d|n}$, $F_i^{d|p} = F_i^{u|n}$) and uses the following values at zero momentum transfer $Q^2=0$:\footnote{Let us note that $F_2^{s|n} \neq 0$ is considered, e.g., in \cite{Bishara:2017pfq}, but $F_2^{s|p}(0) \ll F_2^{u,d|p}(0)$.}
\begin{align}
    & F_1^{u|p}(0) = 2 \, , \quad & & F_1^{d|p}(0) =1\, , \quad  & &F_1^{s|p}(0) = F_1^{s|n}(0)=0 \, ,\\
    & F_2^{u|p}(0) = 2\mu_p + \mu_n -1 \, , \quad  & &F_2^{d|p}(0) = 2\mu_n + \mu_p -1 \, , \quad & &F_2^{s|p}(0) = F_2^{s|n}(0)=0\,.
\end{align}
Here $\mu_p = 2.7930$ and $\mu_n = -1.913042$ represent the proton and neutron anomalous magnetic moments, respectively. The dependence of the form factors on $Q^2$ is given by:
\begin{equation}
    F_1^N (Q^2) = \frac{F_1^N(0) + \tau \left[F_1^N(0) + F_2^N(0)\right]}{(1+\tau)(1+Q^2/M_V^2)^2}\, , \quad  F_2^N (Q^2) = \frac{F_2^N(0)}{(1+\tau)(1+Q^2/M_V^2)^2}
\end{equation}
where $M_V = 0.84\,$GeV. The values of $M_V$, $\mu_p$ and $\mu_n$ correspond to the default values used in the \GENIE module~\cite{Berger:2018urf}.

At low energies, the vector couplings lead to the SI scattering of DM with nuclei. To compare our results with the existing direct detection limits on SI interactions of halo DM, we calculate the quantity $(d\sigma_{\chi p}/d Q^2)\, Q^2_{\max}$ in the highly non-relativistic limit, $Q^2 \to 0$, $s \to (m_\chi + m_p)^2$, that, in this limit, corresponds to the total DM-proton cross section:
\begin{equation}\label{defsigNR}
\sigSI = \frac{g_{Z'}^4 (3Q_q^V)^2 (Q_\chi^V)^2 \mu_{\chi p}^2}{\pi m_{Z'}^4}\,.
\end{equation}
Here $\mu_{\chi p}$ represents the reduced mass of the DM-proton system.

\paragraph{Axial-vector couplings.}
For the charge assignment~\eqref{eq:axial_charges}, $Q_\chi \equiv Q_\chi^A$ in \eq\eqref{dsig_dq2} and
\begin{align}
A&=  (F_A^N - \tau F_P^N)^2 \delta \left(1+\frac{Q^2}{m_{Z'}^2}\right)^2 + (F_A^N)^2 (1+\tau)(\delta + \tau)\, ,\\
C&= (F_A^N)^2\,.
\end{align}
The form factors are now related to quark axial charges:
\begin{equation} \label{sumFA}
    F_i^{N} = \sum_{q=u,d,s} Q_q^A F_i^{q|N}\, ,
\end{equation}
$i=A,P$. The axial form factors are related in isospin limit as  $F_A^{u|p} = F_A^{d|n}$, $F_A^{d|p} = F_A^{u|n}$
 and $F_A^{s|p} = F_A^{s|n}$ and their $Q^2=0$ values are denoted by
 \begin{equation} \label{defFA0}
     F_A^{q|p} (0) = \Delta q\, ,
 \end{equation}
where $\Delta u =0.827$, $\Delta d = -0.38$ and $\Delta s = -0.0427$. 
The dipole form is assumed also for the axial form factor:
\begin{equation}
    F_A (Q^2) = \frac{F_A(0)}{(1+Q^2/M_A^2)^2}
\end{equation}
with $M_A=0.99\,$GeV. 
For the pseudoscalar form factors, Ref.~\cite{Berger:2018urf} assumes $F_P^{s|p}=F_P^{s|n}=0$ and 
\begin{equation}\label{pseudo-scalar_ff}
    F_P^{u/d|p} = \frac{2m_p^2}{{(1+Q^2/M_A^2)^2}} \left( \pm \frac{\Delta u - \Delta d}{m_\pi^2 + Q^2} + \frac{\Delta u + \Delta d - 2 \Delta s}{m_\eta^2 + Q^2}\right)\,
\end{equation}
with $m_\pi = 0.1349766\,$GeV and $m_\eta = 0.547862\,$GeV representing the pion and $\eta$ masses, respectively. The neutron form factors can be again obtained using the isospin relations $F_P^{u|p} = F_P^{d|n}$, $F_P^{d|p} = F_P^{u|n}$. 
Notably, the shapes of the form factors and values of the constants above again correspond to the defaults in the current version of the \GENIE code that may differ slightly from other works, such as from Ref.~\cite{Bishara:2017pfq}.\footnote{In particular, the ChPT treatment~\cite{Bishara:2017pfq} suggests $F_P^{s|p},F_P^{s|n}\neq 0$ and an extra factor of 1/3 in front of the second term in~\eqref{pseudo-scalar_ff} containing the $\eta$ pole. This adjustment would lead to tiny enhancement of the cross section~\eqref{dsig_dq2} in the axial case.}

The axial couplings lead to SD coupling of DM with nuclei at low energies\footnote{In general, axial-vector couplings lead to two different low-energy operators: the standard operator for the SD scattering denoted by $\mathcal{O}_4$ in~\cite{Anand:2013yka,Bishara:2017pfq}, but also the operator $\mathcal{O}_6$ from the same references that is multiplied by the pseudoscalar form factor $F_P$. For our choice of charges $Q^A_u = Q^A_d$, the contribution of the pion pole to $F_P$ vanishes. Hence, $F_P$ becomes very small at low momentum transfers, and considering only the SD contribution $\mathcal{O}_4$ is a good approximation. Overall, the vector couplings lead at leading order solely to the standard SI scattering with nuclei at low energies \cite{Bishara:2017pfq}.} and the formula for the total cross section in the highly non-relativistic limit now reads
\begin{equation}\label{def_sigSD}
    \sigSD = \frac{3 g_{Z'}^4 (Q_\chi^A)^2 \mu_{p\chi}^2 F_A^p(0)^2}{\pi m_{Z'}^4}\, ,
\end{equation}
where $F_A^p (0) $ is the $Q^2=0$ value of the axial form factor, see \eqs\eqref{sumFA} and~\eqref{defFA0}.

\begin{figure}
\begin{center}
\includegraphics[width=0.8\textwidth]{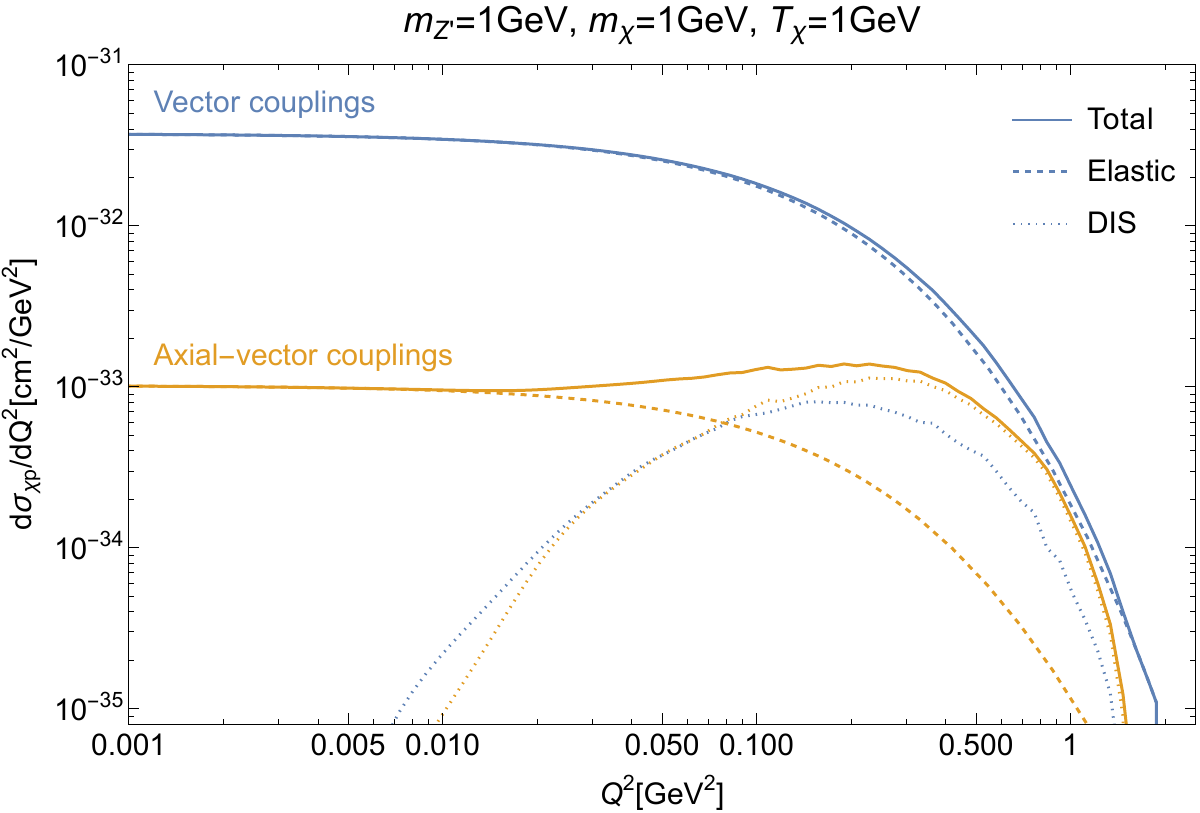}
\end{center}
\caption{Differential cross sections for the scattering of DM with kinetic energy $T_\chi=1\,$GeV with a proton initially at rest for the cases of vector and axial-vector couplings. The mediator and DM masses are set to $\mZp = 1\,$GeV and $\mchi=1\,$GeV, respectively, further, the values $Q_\chi^V = Q_\chi^A = Q_q^V = Q_q^A =1$ for all quarks $q$ and $\gZp=0.1$ are assumed. The dashed lines correspond to the elastic scattering given by formula~\eqref{dsig_dq2}, the dotted lines correspond to the DIS contribution obtained from the \GENIE module~\cite{Berger:2018urf}, and the solid lines correspond to the sum of these two contributions.}
\label{fig:comSig}
\end{figure}

Finally, DIS of DM with nucleons becomes relevant at higher energies and this contribution is calculated numerically using the \GENIE code~\cite{Berger:2018urf}.\footnote{Let us note that \GENIE does not explicitly include the excitation of $\Delta$ resonance and similar higher baryon states. On the other hand, \GENIE output for the DIS cross section covers partly this kinematic range by extrapolating the parton distribution functions to lower $Q^2$ using the Bodek-Yang model~\cite{Bodek:2002ps}.} In \fig\ref{fig:comSig}, the differential cross section $d\sigma_{\chi p}/dQ^2$ is shown, including both elastic scattering and DIS for the case of vector and axial-vector couplings. For the kinetic energies of the incoming DM particles at $T_\chi = 1\,$GeV and with the proton initially at rest, the DIS contribution is already significant at larger momentum transfers and is similar in size for vector and axial-vector couplings. The $Q^2\to 0$ value of the elastic contribution to the differential cross section is, however, suppressed for the axial-vector couplings by the factor
\begin{equation}\label{eq:sup_axial}
    \left.\frac{\frac{d\sigma_{\chi p}^A}{dQ^2}}{\frac{d\sigma_{\chi p}^V}{dQ^2}}\right|_{Q^2 = 0} = \frac{[F_A^p(0)]^2}{(3Q_q^V)^2} \left[1 + \frac{2m_\chi^2}{(m_\chi+T_\chi)^2}\right] \xrightarrow{T_\chi\to 0} \frac{\sigSD}{\sigSI} \simeq 0.05\,,
\end{equation}
where the last equality holds for the specific charges we use for our numerical results: $Q_\chi^V = Q_\chi^A = Q_q^V = Q_q^A =1$ for all quarks $q$.

\subsection{Scattering with nuclei}

Coherent DM-nucleus scattering is the dominant process at lower energies. For the vector couplings, the relevant cross section was calculated in~\cite{Alvey:2022pad}. Under the assumption that the DM-nucleon cross sections in the non-relativistic limit are equal, $\sigma^{\rm SI}_n =\sigSI$, following expression is obtained
\begin{align}
\label{diffsig_full_vector}
\frac{d\sigma_{\chi \A}}{d Q^2}=&\frac{A^2 \sigSI}{\mu_{\chi p}^2 Q^2_\mathrm{max}}
\frac{\mZp^4}{(Q^2+\mZp^2)^2}
\times G_\A^2(Q^2)\\
&
\times\frac{1}{4s}
\left\{
\begin{array}{ll}
m_\chi^2Q^2-Q^2s+(s-\mA^2-m_\chi^2)^2& ~~\mathrm{for~scalar~nuclei,}\\
\frac12 Q^4 -Q^2s+(s-\mA^2-m_\chi^2)^2&  ~~\mathrm{for~fermionic~nuclei,}
\end{array}
\right.\nonumber
\end{align}
for a nucleus with atomic number $\A$ and mass $m_A$. Here $G_\A(Q^2)$ is the nuclear form factor capturing the loss of coherence at large momentum transfers. As in~\cite{Alvey:2022pad}, we employ the model-independent form factors~\cite{Duda:2006uk} based
on elastic electron scattering data. The Fourier-Bessel expansion approach is used,
with parameters taken from \Refe\cite{DeVries:1987atn}. We use model-independent Sum of Gaussian form factors for nuclei where these parameters are not available. Furthermore, $\sigSI$ was defined in \eq\eqref{defsigNR}. For the scattering of DM with kinetic energy $T_\chi$ with a nucleus at rest, the kinematic variables are given by: 
\begin{equation}\label{Q2max_Arest}
    Q^2_{\max} = \frac{4\mA^2}{s}\left(T_\chi^2 + 2\mchi T_\chi\right)\,, 
\end{equation}
and
\begin{equation}\label{s_Arest}
    s = (\mA+\mchi)^2 + 2 \mA T_\chi\,. 
\end{equation}
For the axial couplings, the cross section for coherent elastic scattering vanishes for nuclei with zero spin. Consequently, this process will not be included in this work since the dominant isotopes of larger CR nuclei in our analysis (He, C, O) have zero spin. Additionally, the analysis neglects the nuclei with non-zero spin in our estimate of the attenuation effect, as further described in appendix~\ref{sec:att}.

\begin{figure}
\begin{center}
\includegraphics[width=0.8\textwidth]{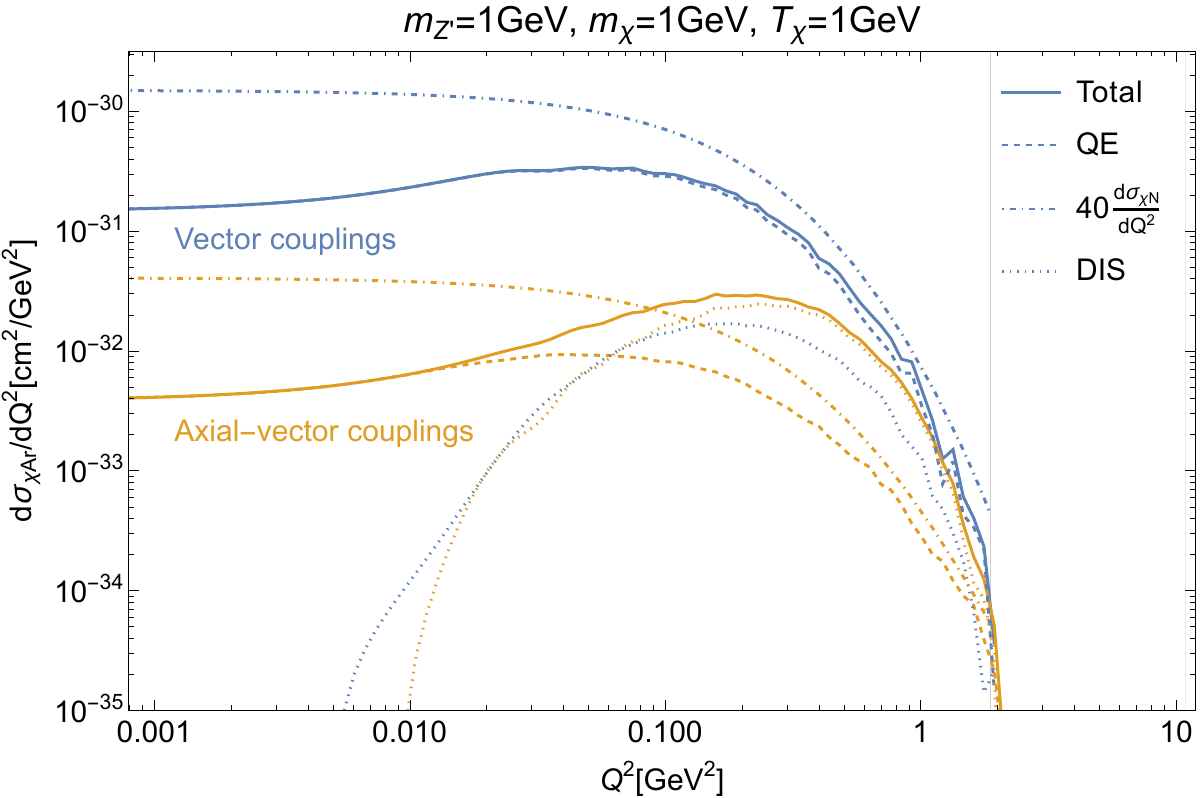}
\end{center}
\caption{Differential cross sections for the scattering of DM with kinetic energy $T_\chi=1\,$GeV with an argon nucleus initially at rest for the cases of vector and axial-vector couplings, obtained from the \GENIE module~\cite{Berger:2018urf}. The values of the couplings and DM and mediator masses are the same as in \fig\ref{fig:comSig}. The dashed and dotted lines correspond to QE scattering and DIS, respectively. The solid lines correspond to the sum of these two contributions. The dot-dashed line depicts the simple estimate of the QE cross section from the sum of elastic cross sections for nucleons initially at rest. The grid lines depict the maximum kinematically allowed momentum transfer $Q^2_{\max}$ for the elastic scattering of DM with a nucleon at rest ($\sim 2\,$GeV$^2$) and with an argon nucleus at rest ($\sim 10\,$GeV$^2$), see appendix~\ref{ap:kinematics} for details.}
\label{fig:comSigAr}
\end{figure}

DM may scatter with individual nucleons within a nucleus through QE scattering at higher energies. 
The full description of this process requires, apart from the knowledge of the DM-nucleon cross section~\eqref{dsig_dq2}, also modeling of the initial momenta of the nucleons and the nuclear potential that affects the final state of the nucleus after the collision. For this reason, the \GENIE module~\cite{Berger:2018urf}, specifically the GDM18\_00a\_00\_000 tune from GENIE v3\_04\_00, is used for calculating the cross sections that are employed for the treatment of DM up-scattering by CR, attenuation, and also DM detection in DUNE, see appendix~\ref{secGENIE} for details. Let us note that in this way, we improve the analysis of \Refe\cite{Ema:2020ulo} that considered the scattering off free nucleons in the DUNE detector, which corresponds to the approximation $\sigma_{\chi A} \sim A \times \sigma_{\chi N}$. We show in \fig\ref{fig:comSigAr} that this simplified formula (depicted by a dot-dashed line) overestimates the cross section for QE scattering of DM off argon nuclei (dashed line).

Finally, DM particles can probe individual partons in the nuclei at the highest energies, corresponding to the DIS process. 
Figure~\ref{fig:comSigAr} shows the relevant DIS cross section as the dotted lines. The DIS cross sections are again comparable for the vector and axial-vector case. In contrast, the QE cross section is suppressed roughly by the factor~\eqref{eq:sup_axial} in the axial case in comparison to the vector case, i.e., we observe roughly the same relative suppression as for the DM-nucleon cross sections depicted in \fig\ref{fig:comSig}.

\section{Results}\label{sec:results}

This section presents the DUNE sensitivities for the two example DM scenarios described above. In the numerical analysis, we choose $Q_\chi^A = Q_\chi^V = Q_q^A = Q_q^V = 1$ $\forall q$, with $\mZp$ fixed at 1\,GeV. Note that with this charge assignment, the DM coupling to protons and neutrons is equal. Our study explores the sensitivity as a function of the DM mass and the coupling constant ($\gZp$) or equivalently the non-relativistic cross sections ($\sigSI=\sigma_n^{\text{SI}}$, $\sigSD=\sigma_n^{\text{SD}}$).

\subsection{CRDM flux}
\begin{figure}
\begin{center}
\includegraphics[width=0.8 \textwidth]{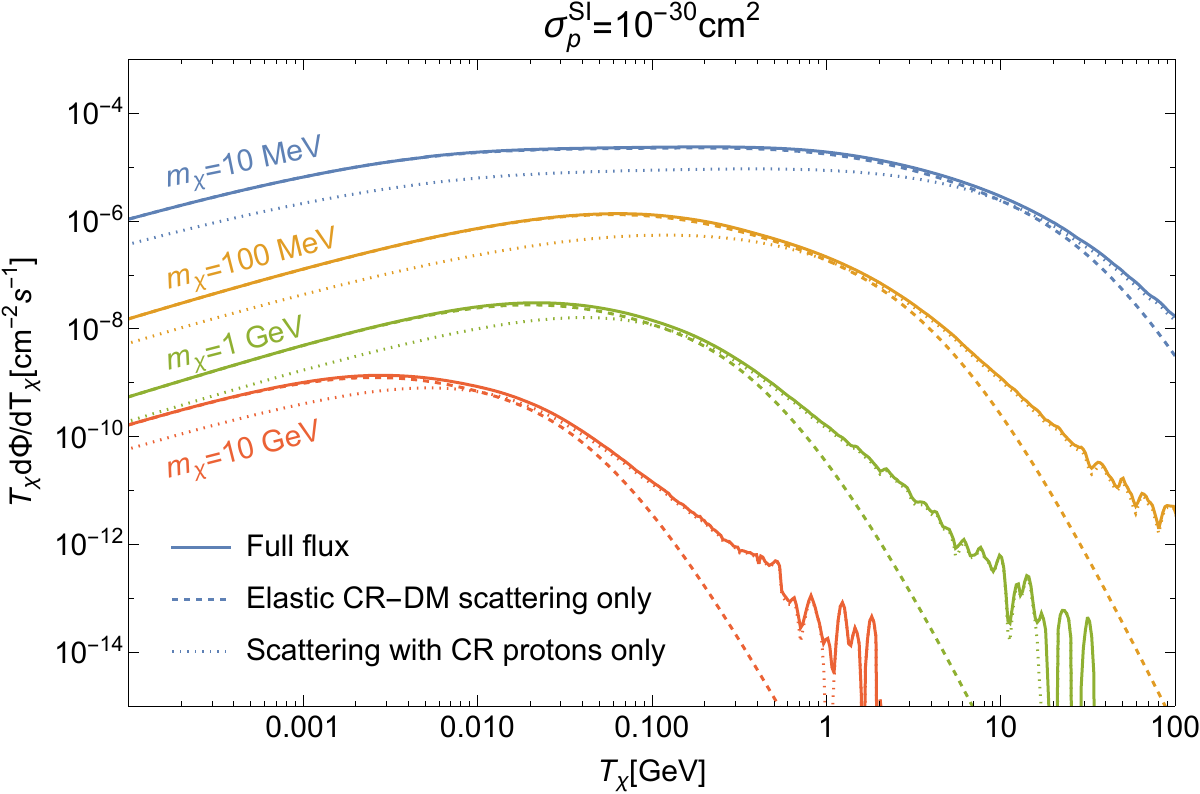} \\
\vspace*{0.3cm}
\includegraphics[width=0.8 \textwidth]{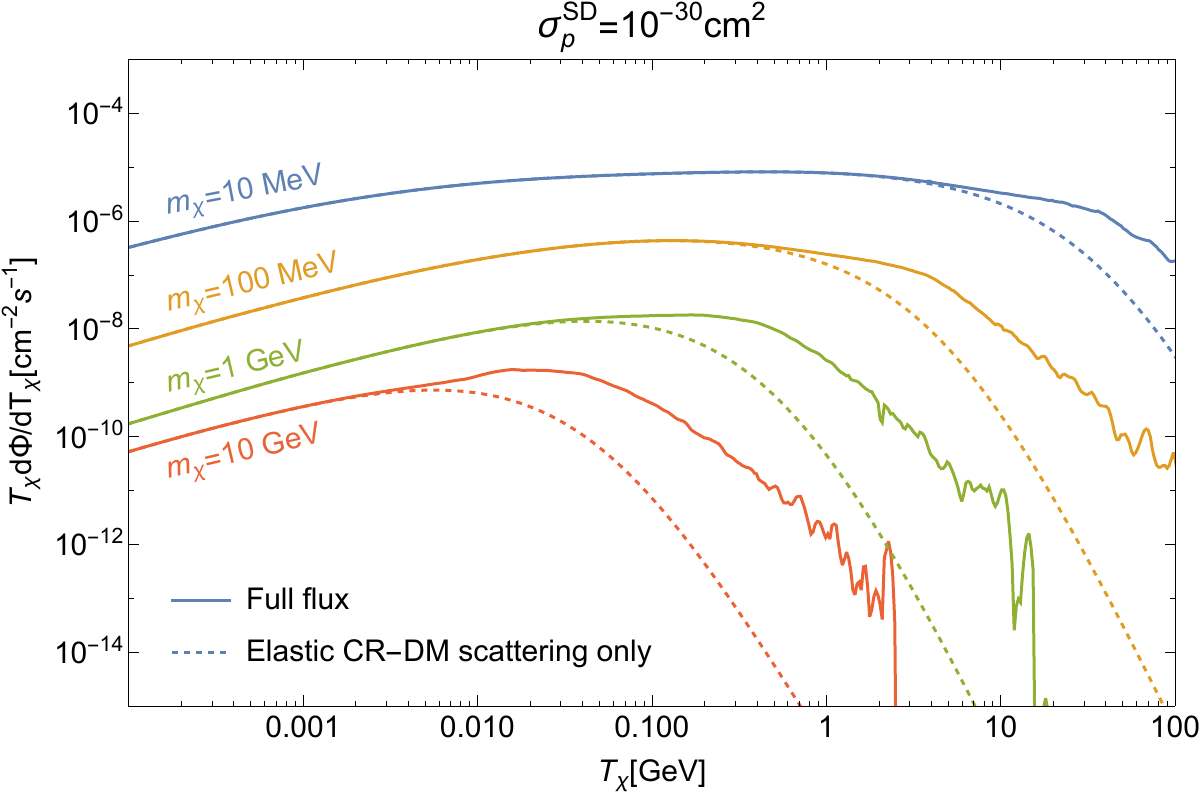}
\end{center}
\caption{CRDM flux coming to the Earth's atmosphere. 
For the vector couplings (top), the solid lines correspond to the flux of the DM particles accelerated by the elastic and inelastic DM scattering with CR protons and He, C, and O nuclei. In contrast, the dotted lines correspond to scattering with protons only. On the other hand, only elastic scattering with the four most abundant CR nuclei is considered for the dashed lines. For the axial-vector couplings (bottom), the solid lines depict scattering with only CR protons (the contribution of the heavier spin-0 nuclei is subdominant in this case). The dashed lines again correspond to the flux from elastic scattering only.} 
\label{fig:flux}
\end{figure}

Using the \ds program~\cite{Gondolo:2004sc,Bringmann:2018lay,Bringmann:2022vra} where the formulas from \sec\ref{sec:CRflux} are implemented, the CRDM fluxes are produced in \fig\ref{fig:flux}. For the vector couplings (top panel), the solid lines depict the full flux due to the scattering of DM with CR protons and He, C, and O nuclei, including both elastic scattering (coherent elastic scattering in case of larger nuclei) and inelastic scattering (DIS for all elements and QE for larger nuclei). The dashed lines then show the case with elastic CR-DM scattering only, whereas the dotted lines correspond to the case where only the scattering with CR protons is included. 

For low DM kinetic energies, the scattering with larger nuclei is important. However, it can be checked that it is mainly the coherent elastic scattering that accounts for the difference between the solid and dotted curves at low energies. For larger $T_\chi$, the contribution of the larger nuclei is negligible due to nuclear form factors that suppress the coherent elastic scattering. By comparison of the solid, dashed, and dotted lines, the contribution of DIS of DM with CR protons significantly increases the flux at large $T_\chi$. 
The tiny difference between the dashed and solid lines around the peak of the CRDM flux is mainly due to QE scattering of DM with CR helium.

Based on the results for vector couplings, only scattering with CR protons is considered for the axial-vector case, where elastic coherent scattering with He, C and O nuclei is absent due to the zero spin of these nuclei. Solid lines in the bottom panel of \fig\ref{fig:flux} depict the fluxes where both elastic scattering and DIS with CR protons are included, whereas the dashed lines include the elastic scattering only. The DIS contribution is even more important for the flux at high $T_\chi$ than for the vector couplings. The higher contribution of DIS at high $T_\chi$ comes from the relative suppression of elastic scattering for the axial-coupling case compared to the vector couplings, as seen in \fig\ref{fig:comSig}.

\subsection{Expected number of events in DUNE}

After obtaining the CRDM flux, the next step calculates the number of events that the DUNE experiment should expect for a given DM model given 400 kiloton*years of data-taking, the baseline exposure for DUNE~\cite{DUNE:2020lwj}. The counting considers a selection cut $\cos \theta_z \geq 0.1$, which corresponds to DM coming from above. Dark matter from below becomes attenuated as it travels through the earth, weakening the signal strength compared to the atmospheric neutrino background. For that reason, the selection cut is used to optimize the signal-to-background ratio. 

\begin{figure}
\includegraphics[width=0.45 \textwidth]{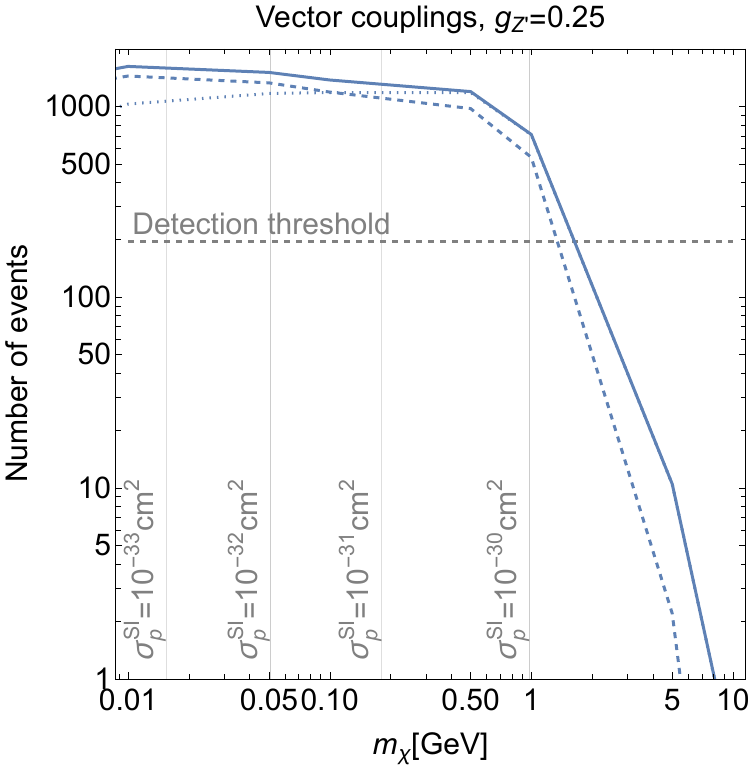} 
\includegraphics[width=0.45 \textwidth]{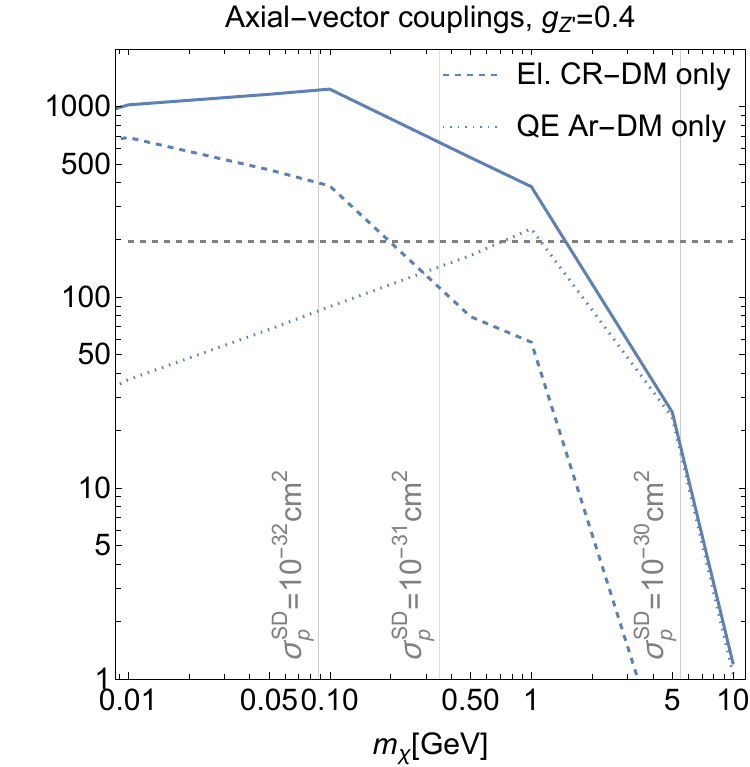}

\caption{Number of events given 400 kiloton*years of data-taking in the signal region $\cos \theta_z \geq 0.1$ possibly seen in DUNE for a fixed value of $g_{Z'}$ and vector (left) or axial-vector (right) couplings. The solid lines correspond to the full CRDM flux induced by both elastic and inelastic scattering with CR and both QE scattering and DIS of DM with argon nuclei in the detector. The dashed lines correspond to elastic-only contribution to the CR-DM scattering, whereas the dotted lines correspond to QE-only scattering with argon. Finally, the dashed gray line corresponds to the detection threshold derived in appendix~\ref{sec:atmnu}.}
\label{fig:events}
\end{figure}

In \fig\ref{fig:events}, solid lines depict the expected total number of events related to DM particles from the signal region $\cos \theta_z \geq 0.1$ for two particular parameter choices. We choose to fix $\gZp$ since then the number of events varies only mildly for smaller DM masses. Grid lines depict the corresponding non-relativistic cross sections. The number of events drops rapidly for $\mchi \gtrsim 1\,$GeV, which is related to the fact that the CRDM flux decreases and contains DM with too small kinetic energies. 
The dashed lines in \fig\ref{fig:events} show the calculated number of DM events when the inelastic interactions are neglected in the CR-DM scattering. As expected based on the observation of the CRDM fluxes, the inclusion of the inelastic CR-DM scattering considerably increases the number of events in the case of axial-vector couplings, especially for larger DM masses. The dotted lines in \fig\ref{fig:events}, in turn, depict the number of events due to full CRDM flux, but when DIS is neglected in the scattering of DM with argon nuclei in the DUNE detector. The DIS contribution is crucial for detecting DM with axial-vector couplings and at lower DM masses. We confirm the findings of~\cite{Kim:2020ipj} showing that DIS becomes less important for DM detection in DUNE-like detectors for DM masses $\gtrsim 1\,$GeV.  Finally, the gray dashed line shows for comparison the threshold number of DM events $\Nthr = 196.7$ that can be distinguished from the atmospheric neutrino background. For the chosen $\gZp$ couplings in \fig\ref{fig:events}, the CRDM particles in the signal region $\cos\theta_z \geq 0.1$ are not considerably slowed down by the Earth's crust or atmosphere. Hence, the number of DM events is calculated directly based on the CRDM fluxes depicted in \fig\ref{fig:flux}.

\subsection{Constraints on dark matter parameter space}

\begin{figure}
\begin{center}
\includegraphics[width=0.8 \textwidth]{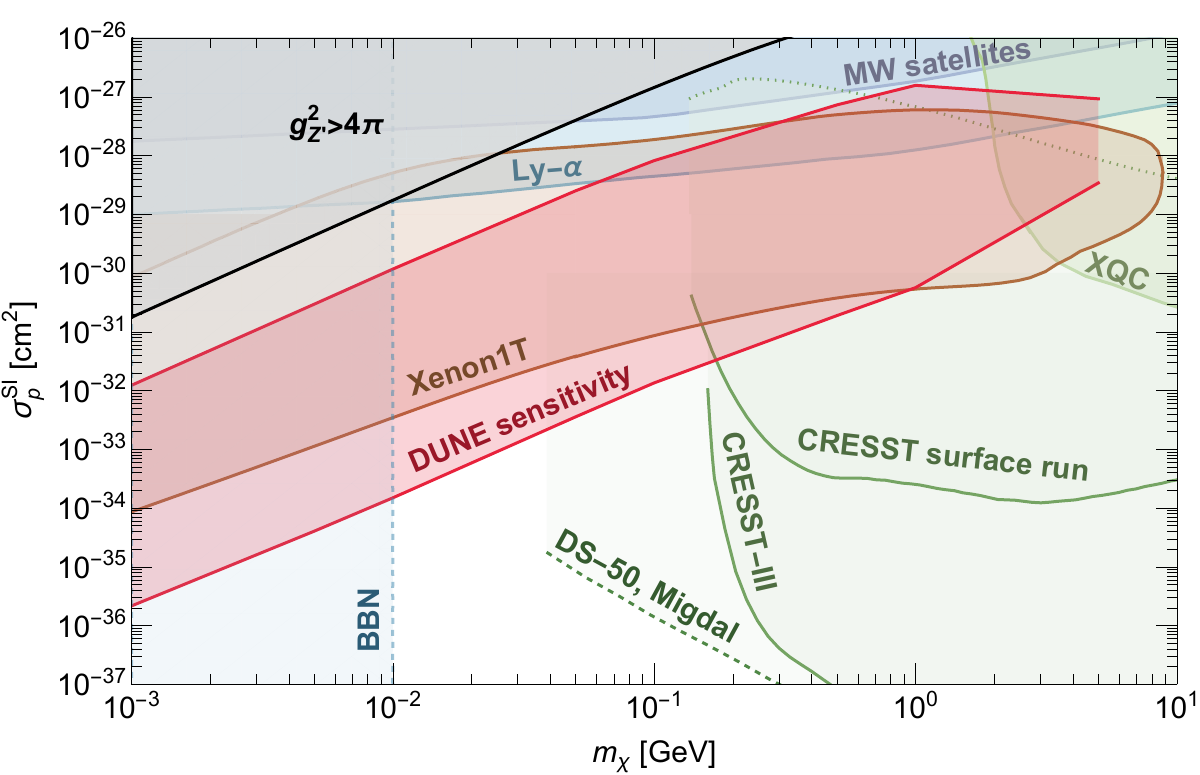} \\
\vspace*{0.3cm}
\includegraphics[width=0.8 \textwidth]{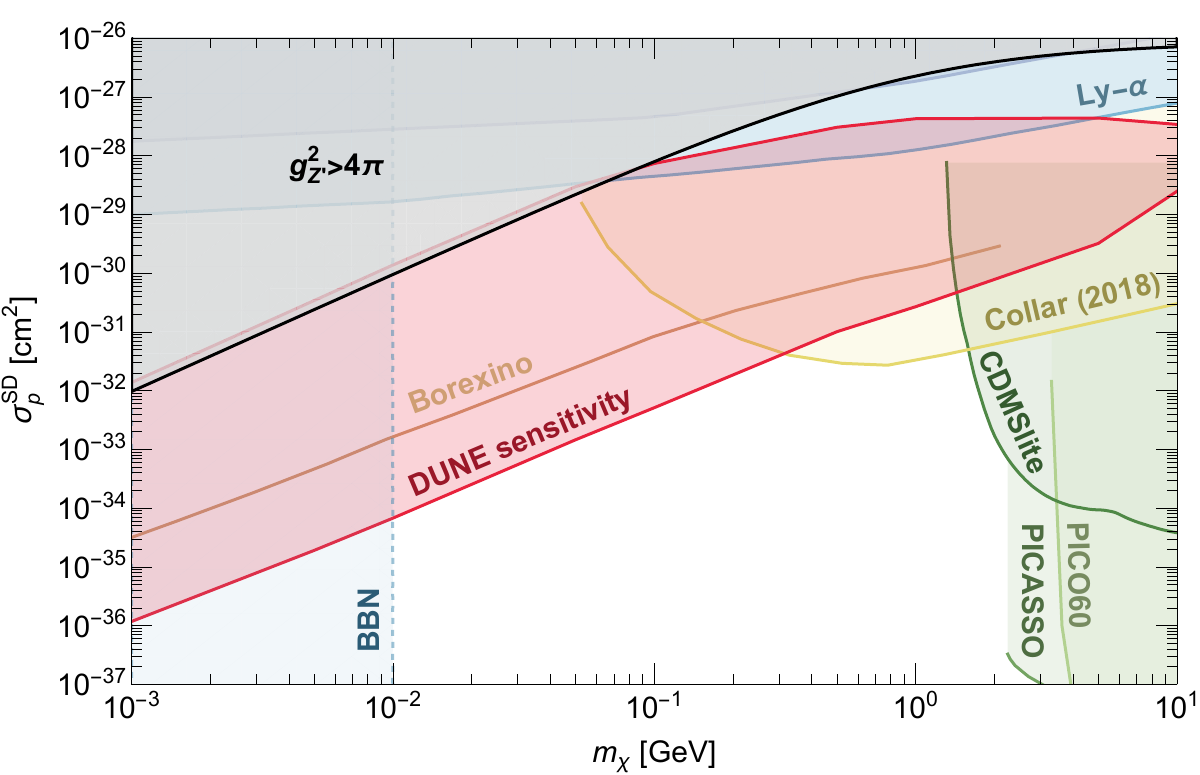}
\end{center}
\caption{Sensitivity of the DUNE detector assuming 400 kiloton*years of data taking and considering only statistical errors on the number of background neutrino events (red region). Other colored regions correspond to complementary constraints. In particular, the brown regions depict the CRDM bounds by Xenon1T~\cite{Diurba:2024dqo} and Borexino~\cite{Dent:2019krz} for similar mediator scenarios as considered in our work. Further, different shades of green depict bounds on halo DM by direct detection experiments while shades of blue are used for cosmological constraints, see the main text for details. The gray region correspond to parameter space where our DM scenarios would become non-perturbative.} 
\label{fig:ExclSig}
\end{figure}

We will now present the sensitivity regions for DUNE given 400 kiloton*years of data-taking. These regions correspond to the parameter space of the different DM scenarios where the predicted number of events is larger than $\Nthr$. In \fig\ref{fig:ExclSig}, we show the sensitivity regions in the $m_\chi - \sigma_p$ plane to allow comparisons to other existing constraints.\footnote{For $\mZp = 1\,$GeV, the typical transferred momentum $Q^2$ in the scattering of halo DM in direct detection experiments satisfies $Q^2 \ll \mZp^2$. Consequently, the results of these experiments can be reinterpreted as results for cross sections in the $Q^2 \to 0$ limit, $\sigSI$ and $\sigSD$. For lighter mediators, rescaling traditional direct detection results would be needed, as shown in Ref.~\cite{Alvey:2022pad}.}

For the vector couplings (top panel of \fig\ref{fig:ExclSig}), corresponding to SI scattering with nuclei, the results are compared with the bounds on CRDM~\cite{Alvey:2022pad, Diurba:2024dqo} obtained by reinterpretation of Xenon1T data~\cite{XENON:2018voc}.\footnote{We choose to present the results of \cite{Diurba:2024dqo} where the inelastic scattering of CR with DM is included, however, the attenuation of the CRDM flux in the Earth's crust is treated in a simplified manner as found in Ref.~\cite{Alvey:2022pad}. The full analysis of the attenuation, including the \GENIE data on the cross sections of DM with nucleons in the soil, roughly confirms the approximate method~\cite{Diurba:2024dqo}.} Notably, the CRDM constraints based on LZ results~\cite{LZ:2025iaw} are stronger than those based on the Xenon1T experiment; however, the results of~\cite{LZ:2025iaw} assume the simplified constant (i.e., $Q^2$-independent) cross section and need to be recast for the case of vector mediators to make comparisons possible. 
Ref.~\cite{Bell:2023sdq} suggests that a limit by LZ is expected to be about a factor of two stronger than the one by Xenon1T, but detailed analysis of the LZ bound is left for future work. Further, we include in \fig\ref{fig:ExclSig} the bounds on nuclear recoils produced by halo DM by the CRESST program (their surface run~\cite{CRESST:2017ues} and the underground CRESST-III run~\cite{CRESST:2019jnq}) and the rocket-based XQC experiment~\cite{McCammon:2002gb,Mahdawi:2018euy}. The DarkSide-50 results~\cite{DarkSide:2018bpj} place stronger constraints than CRESSST-III on DM with masses larger than 1.8\,GeV, however, this would not appear in the cross-section range depicted in our \fig\ref{fig:ExclSig}. Further, strong bounds on SI scattering of light DM have been obtained based on the searches for Migdal electrons produced in nuclear recoils~\cite{XENON:2019zpr,DarkSide:2022dhx}. We depict the DarkSide-50 Migdal constraints~\cite{DarkSide:2022dhx} by a dashed line in our comparison plot due to uncertainties in the modeling of the Migdal effect~\cite{Xu:2023wev}. Let us note that similarly as CRDM, also the halo DM can be stopped by Earth's crust or atmosphere before reaching the detectors, consequently, the direct detection limits do not reach arbitrarily high cross sections. For the CRESST surface run, the corresponding maximal cross section is depicted by a dotted green line obtained from~\cite{Emken:2018run}. For other direct detection experiments, similar analysis of the maximal probed cross section is not present in the existing literature to the best of our knowledge. For the experiments located in the Gran Sasso laboratory (CRESST-III and DarkSide), value of about $10^{-30}\,$cm$^2$ for $\mchi \sim 1\,$GeV might be expected based on the results of~\cite{Emken:2018run} for CRESST-II. 


Finally, we present the limits coming from structure formation, in particular, from the Lyman-$\alpha$ forest~\cite{Rogers:2021byl}, and the Milky Way satellite population~\cite{Maamari:2020aqz}, and also the bounds related to the BBN~\cite{Krnjaic:2019dzc}. The latter bound is related to the fact that for large DM-nucleon cross sections, DM particles would be in thermal equilibrium with SM plasma in the early Universe and if $m_\chi \lesssim 10\,$MeV, they would contribute to the effective number of relativistic degrees of freedom, which would alter the BBN predictions. Let us, however, note that this bound might be slightly loosened depending on the couplings to other SM fermions~\cite{Ema:2020ulo, Escudero:2018mvt,Sabti:2019mhn}, for this reason, we depict the BBN bound by a dashed line. 


In the case of SD scattering (bottom panel of \fig\ref{fig:ExclSig}), \Refes\cite{Bringmann:2018cvk, Dent:2019krz} report constraints on CRDM by the neutrino experiment Borexino~\cite{Borexino:2009mcw,Borexino:2013bot}. While \Refe\cite{Bringmann:2018cvk} does not consider the $Q^2$-dependence of the cross section, \Refe\cite{Dent:2019krz} introduces a mediator with axial-vector couplings similar to this work. Hence, we display only the latter results in \fig\ref{fig:ExclSig}.\footnote{
Ref.~\cite{Dent:2019krz} introduces axial-vector couplings of the mediator directly to nucleons (not quarks). The contribution of the pseudo-scalar form factor to the DM-nucleon scattering is neglected. Further, when presenting their results on the excluded values of $\sigSD$, they fix the gauge coupling and vary the mediator mass, whereas we fix $\mZp$ and vary $\gZp$. On the other hand, it can be checked that also \Refe\cite{Dent:2019krz} constraints $\mathcal{O}$(GeV) mediator masses, so the results should be comparable.} Further complementary limits come from direct detection experiments like CDMS~\cite{SuperCDMS:2017nns}, 
PICO-60~\cite{PICO:2017tgi}\footnote{PICO-60 update from 2019~\cite{PICO:2019vsc} refers to DM cross sections smaller than $\sim 10^{-36}\,$cm$^2$, so it would not bring further excluded parameter space in our comparison plot.} or PICASSO~\cite{Behnke:2016lsk},
and from delayed-coincidence searches in near-surface detectors~\cite{Collar:2018ydf}. Also for these experiments, DM can be stopped by the Earth's atmosphere or crust before reaching the detectors, so the corresponding bounds do not extend to arbitrarily large cross sections. Reach to larger cross sections is expected for the case of~\cite{Collar:2018ydf} based on near-surface detectors than for the underground experiments CDMS, PICO and PICASSO, dedicated analysis would be needed for further details.

\begin{figure}
\begin{center}
\includegraphics[width=0.8\textwidth]{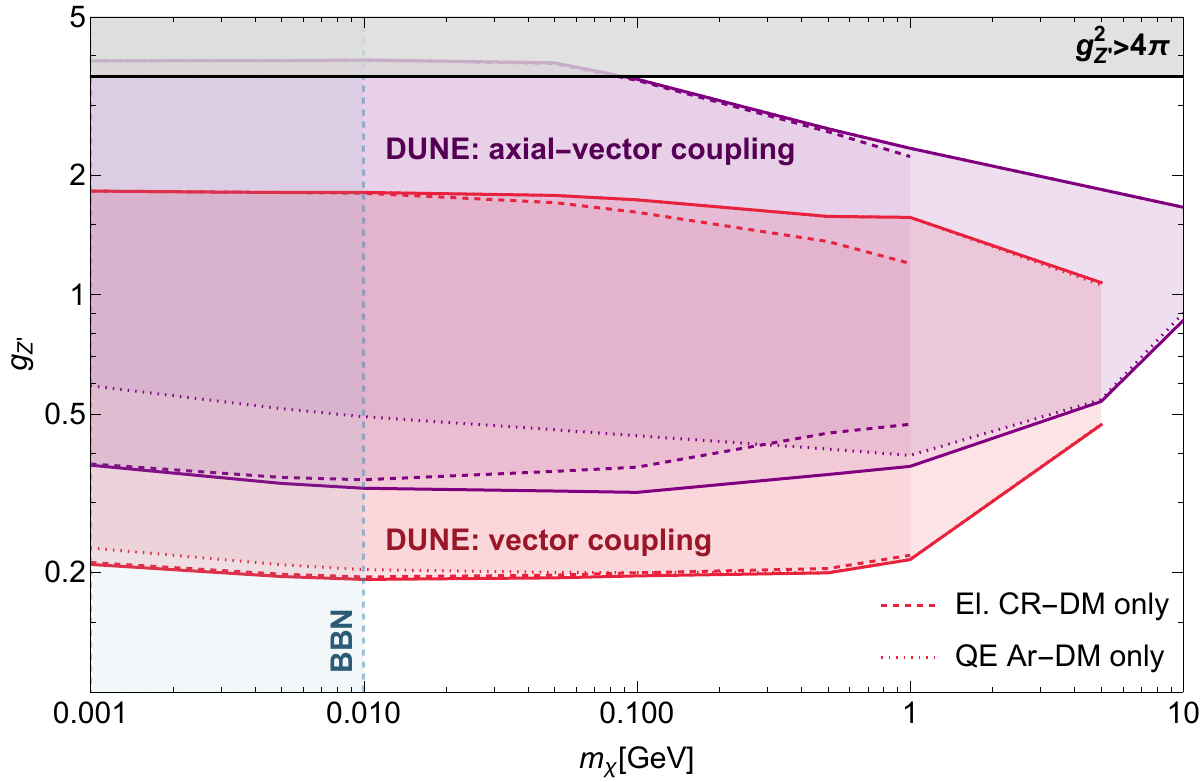}
\end{center}
\caption{Comparison of the values of $\gZp$ that DUNE will be sensitive to for the vector and axial-vector couplings. As in \fig\ref{fig:events}, dashed lines show how the boundary of the sensitivity region is changed if only elastic scattering of DM with CR is included in the analysis. The dotted lines, in turn, correspond to the case of QE-only detection in DUNE.}
\label{fig:Exclg}
\end{figure}

In \fig\ref{fig:Exclg}, we compare the DUNE sensitivity region in the $\mchi-\gZp$ plane obtained for vector and axial-vector couplings. Using the same type of lines as in \fig\ref{fig:events}, we also show how the sensitivity region shrinks when only elastic scattering of DM with CR is included (dashed lines) and when only QE scattering of DM with argon nuclei in the DUNE detector is considered. The suppression of the DM-nucleon cross section for the axial-vector couplings compared to the vector ones (see \fig\ref{fig:comSig}) leads to weaker limits on $g_{Z'}$ in the axial-vector case. It also makes the contribution from DIS stronger in the axial-vector case. The inelastic scattering of DM with CR is important for the largest DM masses, without this contribution, DUNE would not be able to probe DM masses of $\mchi = 5\,$GeV neither in the vector nor axial-vector case.\footnote{Since the inelastic cross sections need to be generated by the \GENIE code separately for each DM mass and this is computationally costly, results are evaluated only in the discrete points $m_\chi/\text{GeV}=0.001,\,0.005,\,0.01,\,0.05,\,0.1,\,0.5,\,1,\,5,\,10$.}




\subsection{Discussion on other datasets}

Comparison to previous works that presented CRDM constraints by neutrino experiments~\cite{Ema:2020ulo,Super-Kamiokande:2022ncz} is not provided directly in \fig\ref{fig:ExclSig} since these results used different models, namely, scalar or pseudoscalar mediators. For the pseudoscalar mediator case, it is impossible to compare to our results for DUNE since the corresponding interaction of DM with quarks does not lead to SD scattering with nuclei at low energies, as in the case of axial-vector couplings. On the other hand, based on the results of~\cite{Alvey:2022pad}, we expect that the sensitivity regions in terms of the non-relativistic cross sections should not differ considerably for the case of vector and scalar mediators, and our work indeed obtained DUNE sensitivities for similar $\sigSI$, as in~\cite{Ema:2020ulo}. 

Ref.~\cite{Ema:2020ulo} further reports on the DUNE sensitivity region in terms of the product of the scalar mediator coupling to quarks and to DM $g_{\phi \chi} g_{\phi q}$ that is analogous to the quantity $\gZp^2 Q_\chi^V Q_q^V$ in our work. One can see in~\cite{Ema:2020ulo} that considerably smaller values of $g_{\phi \chi} g_{\phi q}$ can be probed by DUNE compared to the values of $\gZp^2$ that can be inferred from \fig\ref{fig:Exclg}. This difference occurs because the coupling of a scalar mediator to nucleons is strongly enhanced compared to its coupling to quarks: $g_{\phi N} \sim \sqrt{300} g_{\phi q}$, while in the case of vector mediator, the coupling of $Z'$ to nucleons is of the same order as the coupling to quarks at $g_{Z' N} = 3 Q_q^V \gZp$.

Moreover, \Refe\cite{Ema:2020ulo} derives bounds based on KamLAND~\cite{KamLAND:2011fld} and SuperK~\cite{Super-Kamiokande:2009kfy} data and predicts sensitivities of future detectors JUNO~\cite{JUNO:2021vlw} and HyperK~\cite{Hyper-Kamiokande:2018ofw}. The results of~\cite{Ema:2020ulo} for both the scalar and pseudoscalar mediator show that these experiments do not probe as small DM couplings as DUNE. We expect that also in the case of vector and axial-vector couplings discussed in our work, the DUNE experiment will be most sensitive, consequently, we do not consider other experiments in detail. Despite its large volume, Hyper-K is expected to be less sensitive to CRDM due to larger threshold for proton detection (\Refe\cite{Ema:2020ulo} assumed the detection threshold of 485\,MeV for Hyper-K while for DUNE the kinetic energy necessary for detection is 40\,MeV). Further, the sensitivity of the liquid scintillator detector JUNO (similarly to the existing limits by KamLAND) was derived in~\cite{Cappiello:2019qsw,Ema:2020ulo,Chauhan:2021fzu} by considering the signal from elastic scattering with hydrogen, the signal from elastic scattering with carbon nuclei is expected to be strongly quenched~\cite{Dasgupta:2011wg,Cappiello:2019qsw,Chauhan:2021fzu}. Consequently, only part of the detector mass was assumed to form targets for DM detection and, therefore, weaker bounds were obtained. Using scattering with hydrogen only, we expect the JUNO sensitivity region to be smaller than for DUNE also in our case.\footnote{Let us note that energetic DM may in principle contribute to scattering with nucleons within carbon nuclei and this process might be detected in JUNO similarly as atmospheric neutrinos~\cite{Settanta:2019ecp}. However, a detailed analysis of the signal in JUNO is beyond the scope of this work.} On the other hand, all the planned large underground neutrino experiments have the potential to probe the unexplored DM parameter space~\cite{Ema:2020ulo} and in the presence of a DM signal, combining the data from all these experiments located in different parts of the world could provide further insights.




Finally, let us note that we compared the CRDM constraints mainly to standard direct detection experiments that depend on the non-relativistic DM-nucleon interactions only. 
Refs.~\cite{Elor:2021swj,Bell:2023sdq} suggest that the scenarios with sub-GeV mediators might be strongly constrained by other complementary probes like meson decays, however, these probes are model-dependent, in particular, depend on the $Z'$ coupling to heavier quarks that are not important for our study. Consequently, a discussion of further theoretical and experimental constraints on the $Z'$ scenarios is beyond the scope of this work. An analogous study is performed in~\cite{Ema:2020ulo} for the case of scalar and pseudoscalar mediators and the results suggest that when CRDM is taken into account, DUNE might be sensitive to DM parameter space not yet constrained by any other experiment.

\section{Conclusions and outlook}\label{sec:conc}

This work presents the sensitivity of the planned DUNE experiment to sub-GeV DM particles accelerated by scattering with galactic CR. The DM interactions occur with ordinary matter via a new gauge boson $Z'$ that couples to SM quarks and DM either through vector or axial-vector interactions. The latter scenario is much more weakly constrained by standard direct detection experiments, as it corresponds to SD scattering with nuclei. However, we demonstrated that the CRDM constraints are comparable in both cases. This result points to an essential feature of the CRDM constraints, namely, that their presence is inevitable for any model where DM couples to nucleons since both the acceleration and detection mechanisms rely on this coupling. The results encourage further study of the CRDM constraints for other scenarios that might be less constrained by standard direct detection experiments, such as those where the low-energy interactions of DM with nucleons are velocity-suppressed.

Our work improved the previous studies by more realistic modeling of the QE scattering of DM with nuclei thanks to the \GENIE module~\cite{Berger:2018urf} and also by the inclusion of DIS process both for the DM up-scattering and detection. We observe that the DIS process is vital for the axial-vector couplings where the elastic scattering is relatively suppressed compared to the vector case (see \fig\ref{fig:comSig}). More detailed comparisons of the two panels in \fig\ref{fig:ExclSig} reveal that the bounds on the DM-nucleon non-relativistic cross section are by about a factor of two stronger in the axial-vector case than for the vector couplings, mainly due to the DIS contributions.

We also improved on previous works by considering the atmospheric neutrino background relevant for the possible DM detection in DUNE. Let us note that a more efficient background subtraction would be possible if the directionality of the signal is taken into account, in particular, the CRDM flux is expected to arrive predominantly from the direction of the galactic center where the most significant DM density is expected~\cite{Ge:2020yuf, Super-Kamiokande:2022ncz, Nagao:2022azp, NEWSdm:2023qyb}. Study of the resulting anisotropy of the CRDM flux requires careful modeling of both the DM and CR distributions, and we believe this is another interesting avenue for future work. 

On the topic of modeling particle scattering, another future path would be comparing the different outgoing particles from DM and neutral current neutrino interactions. Backgrounds can be rejected by selecting interactions based on the number and energies of outgoing particles predicted for DM interactions. These backgrounds also have neutrino scattering modeling uncertainties that impact the systematic uncertainties for this CRDM search, which are outside the scope of this work. We are excited at the prospect of existing and future neutrino cross sections from the Short Baseline Neutrino (SBN) program~\cite{MicroBooNE:2015bmn}. Investigating data from the SBN program allows the community to build thorough models with constrained uncertainties for this neutral current background, allowing us to include informed modeling uncertainties in future sensitivity studies.

\section*{Acknowledgments}
We thank Torsten Bringmann for useful discussions, coding tips and comments on the manuscript. We further thank Josh Berger and Martin Hoferichter for helpful discussions, 
and to James Alvey, Gonzalo Herrera, and Gailyn Monroe for useful input in the initial stages of this project. Finally, we thank to AEC, University of Bern for providing an environment that allowed for starting this collaboration between the two authors. H.K. is supported by the Research Council of Norway under the FRIPRO Young Research Talent grant no.~335388.

\appendix
\section*{Appendix}
\section{Kinematics of dark matter scattering with nucleons and nuclei}\label{ap:kinematics}
When calculating the CRDM flux~\eqref{eq:CRflux1}, integration bounds for the CR kinetic energy must be determined. It is, therefore, necessary to study the ranges that different kinematic variables can attain for different scattering processes. 

Let us first consider the scattering of a DM particle with a nucleon, the initial 4-momenta of these particles being $k = (E_\chi, \vec{k})$ and $p$, respectively. The (square of the) invariant mass of this system is then defined as $s=(k+p)^2$. Let us further denote the final DM 4-momentum by $k' = (E_\chi',\vec{k'})$ and the 4-momentum of the final hadronic system by $W$, where $\mN^2 \leq W^2 \leq \sqrt{s} - \mchi$, with the lower and upper bounds for $W^2$ corresponding to the elastic collision and the perfectly inelastic collision, respectively. The 4-momentum exchange $Q^2 \equiv -q^2 \equiv - (k-k')^2$ can be then written as
\begin{equation}\label{eq:Q2}
    Q^2 = 2\left( E_\chi E_\chi' - \mchi^2 - \sqrt{E_\chi^2 - \mchi^2}\sqrt{E_\chi'^2 - \mchi^2} \cos \theta\right)
\end{equation}
where $\theta$ is the angle between the vectors $\vec{k}$ and $\vec{k'}$ and this relation holds in any reference frame. In order to find the possible range of $Q^2$, it is convenient to use the center-of-mass frame where
\begin{equation}
    E_\chi^\cm = \frac{s+\mchi^2 - \mN^2}{2\sqrt{s}}\,\qquad E_\chi'^\cm = \frac{s+\mchi^2 - W^2}{2\sqrt{s}}.
\end{equation}
The maximum value of $Q^2$ is then obtained for $\cos\theta_\cm = -1$ and for the minimum value of $W^2$: $W^2_{\min} = \mN^2$. Consequently, maximum $Q^2$ is achieved in an elastic collision where \eq\eqref{eq:Q2} can be simplified and
\begin{equation}\label{Q2maxGen}
    Q^2_{\max} = \frac{\left(s+\mchi^2-\mN^2\right)^2}{s}-4 \mchi^2\,.
\end{equation}
For the case of DM with kinetic energy $T_\chi$ scattering with nucleons initially at rest, Mandelstam $s$ and $Q^2_{\max}$ are given by formulas~\eqref{eq:s_Nrest} and~\eqref{Q2max_Nrest}, respectively.

For scattering with nuclei initially at rest, we need to distinguish between different possible processes. At low energies, coherent scattering with the whole nucleus is dominant, and the maximum $Q^2$ for this process is given by \eq\eqref{Q2max_Arest}. At higher energies, scattering with individual nucleons becomes relevant, formula~\eqref{Q2maxGen} then applies, but $Q^2_{\max}$ might be larger than in formula~\eqref{Q2max_Nrest} since the nucleons are not initially at rest, they are moving within the nucleus, the typical momenta being $\mathcal{O}(10)-\mathcal{O}(100)\,$MeV, depending on the size of the nucleus. The numerical output from the \GENIE code, however, suggests that $Q^2$ in the inelastic scattering of DM with nuclei is smaller than $Q^2_{\max}$ for the coherent elastic scattering, see \fig\ref{fig:comSigAr} for the example of DM-Ar scattering. In this figure, the grid line at $Q^2 \sim 2\,$GeV$^2$ depicts the value of $Q^2_{\max}$ for the elastic scattering of DM with kinetic energy $T_\chi = 1\,$GeV with a nucleon at rest and we see that the cross section indeed quickly drops for larger $Q^2$. In particular, non-negligible values of the cross section do not reach the second grid line at $Q^2 \sim 10\,$GeV$^2$ that corresponds to $Q^2_{\max}$ for the coherent elastic DM-Ar scattering.

In summary, for DM with fixed kinetic energy scattering off a nucleus initially at rest, the largest $Q^2$ is obtained for coherent elastic scattering. Using this information, the value of $T_\A^{\min}$ entering \eq\eqref{eq:CRflux1} can be found in the following way. First, note that for a given nucleus $\A$, the DM kinetic energy in the nucleus rest frame~\eqref{eq:tildeTchi} is fixed by $T_\A$. Consequently, for fixed $T_\A$, the largest $Q^2$ is again obtained for coherent elastic scattering. For this process, the kinematic variables can then be evaluated in the DM rest frame that is more suitable for the study of the CR-DM scattering. The Mandelstam $s$ for the case of a CR nucleus $\A$ scattering with DM particles at rest is analogous to \eq\eqref{s_Arest}, however, with the role of $\A$ and $\chi$ exchanged: 
\begin{equation}
    s = (\mA+\mchi)^2 + 2 \mchi T_\A \,.
\end{equation}
Using $Q^2_{\max} = 2\mchi T_\chi^{\max}$, we obtain the maximum kinetic energy of the up-scattered DM particle
\begin{equation}
      T_\chi^{\max}  =   \frac{2\mchi^2}{s}\left(T_\A^2 + 2\mA T_\A\right)\,.
\end{equation}
From this relation, \eq\eqref{eq:TAmin} is obtained.

Finally, as mentioned above \eq\eqref{eq:doublediff}, two kinematic variables are needed to describe the inelastic scattering of DM with nuclei or nucleons. The 4-momentum transfer $Q^2$, the angle between incoming and outgoing DM particles $\theta$ and the hadronic invariant mass $W$ were already mentioned above. Let us add for completeness that other kinematic variables might be used, such as the Lorenz invariant Bjorken $x$, or inelasticity $y$.

\section{Attenuation of the CRDM flux}\label{sec:att}

As explained in \sec\ref{sec:CRflux}, part of the CRDM particles can be stopped by the Earth's atmosphere or crust. For the simplified modeling of this effect, an energy-loss equation for the DM kinetic energy at a depth $z$ is used:
\begin{equation}
    \label{eq:eloss}
\frac{dT_\chi^z}{dz}=-\sum_\A n_\A \int_0^{\omega_\chi^\mathrm{max}}\!\!\!d\omega_\chi\,\frac{d \sigma_{\chi \A}}{d\omega_\chi} \omega_\chi\,.
\end{equation}
It was shown in~\cite{Emken:2018run} that this approach typically overestimates the attenuation effect compared to the Monte-Carlo techniques. Hence, our evaluations should be conservative in this sense. In \eq\eqref{eq:eloss}, $\omega_\chi \equiv T_\chi - T_\chi'$ is the energy loss of the DM particle in a single collision with a nucleus $\A$ initially at rest. Let us note that for elastic scattering, $\omega_\chi = Q^2/(2\mA) = T_\A$ where $T_\A$ is the kinetic energy of the recoiled nucleus. Importantly, for large $Q^2$, the cross section of the coherent elastic scattering is suppressed by nuclear form factors. Consequently, without the inclusion of inelastic scattering, the attenuation effect is substantially underestimated~\cite{Alvey:2022pad,Su:2022wpj}. In this work, the inelastic scattering of DM with nuclei and the relevant contribution to~$d \sigma_{\chi \A}/d\omega_\chi$ comes from the \GENIE code~\cite{Berger:2018urf}. In principle, all the DM kinetic energy can be lost in the collisions from inelastic processes, hence, $\omega_\chi^{\max} = T_\chi$ in formula~\eqref{eq:eloss}.

In the case of underground detectors like DUNE, the attenuation is dominated by the contribution of the Earth's crust. Consequently, the sum in \eq\eqref{eq:eloss} runs over the nuclei in the Homestake rock, with $n_\A$ being the corresponding number densities. These are calculated from the soil density (assumed to be equal to $2.7\,$g/cm$^3$ that was used in~\cite{Ema:2020ulo} and that lies in the density range reported in~\cite{DensHomestake}) and the mass fractions of different elements. Inspired by Homestake geology~\cite{GeoHomestake}, this work assumes 52.7\% of oxygen, 31.5\% of silicon, 6.2\% of aluminum, 3.9\% of magnesium, 3.8\% of calcium and 1.9\% of potassium. 
These fractions correspond roughly to the average among the different rock formations; however, the final results are not sensitive to their variation. Among these nuclei, only potassium and aluminum have dominant isotopes with non-zero spin, and could then, in principle, contribute to the coherent elastic scattering in the case of axial couplings that lead to the SD scattering with nuclei. On the other hand, the coherent elastic scattering is not enhanced by the factor of $A^2$ in the SD case and, in general, this process predominantly influences DM with lower energies that may not pass the detection threshold in DUNE. Consequently, we neglect the coherent elastic scattering in the case of axial couplings.


As mentioned in the main text, the signal region is restricted to the zenith angles with $\cos\theta_z \geq 0.1$, and the number of events in the DUNE detector is then not affected by the attenuation effect for DM with masses $\lesssim 1\,$GeV and the cross sections at the lower boundary of the DUNE sensitivity region. However, the CRDM flux becomes attenuated for larger cross sections and masses, and this is taken into account when calculating the event rate~\eqref{eq:rateDUNE}. 
In practice, the flux~\eqref{eq:CRflux2} is split into 20 parts related to different length intervals of 0.1 in $\cos\theta_z$. Only nine of these parts then contribute, given the cut of $\cos\theta_z \geq 0.1$. For each part, the analysis evaluates the attenuation by the soil layer with the largest thickness within the given interval of $\cos\theta_z$, assuming Earth to be approximately spherical. For example, for the DM particles coming from the directions with $0.1\leq \cos\theta_z \leq 0.2$, the soil layer is quantified to be $14.8\,$km. Consequently, integration over $\cos\theta_z$ in \eq\eqref{eq:rateDUNE} is performed as a sum over the nine bins in $\cos\theta_z$ mentioned above. Let us note that the differential equation~\eqref{eq:eloss} is solved using the \ds code~\cite{Gondolo:2004sc,Bringmann:2018lay,Bringmann:2022vra}, again supplemented by the inelastic DM-nucleus cross sections obtained from \GENIE~\cite{Berger:2018urf}.


\section{Atmospheric neutrino background}\label{sec:atmnu}

The signatures of the boosted DM in DUNE might be very similar to the ones of neutral-current interactions of atmospheric neutrinos. To estimate the related background, we consider the atmospheric neutrino flux presented in the works by Honda et al.~\cite{Honda:2006qj, Honda:2011nf, Honda:2015fha}. The corresponding data are available at the web page~\cite{HondaWeb}, in particular, we use the tables 
for averaged fluxes  
$d\overline{\Phi}_{\nu}^j/(dE_\nu\,d\Omega)$ in the units of  (m$^2$\,s\,sr\,GeV)$^{-1}$ describing the neutrinos arriving to the Homestake location from the zenith angles $0.1\, j \leq \cos \theta_z\leq  0.1\, (j+1)$, $j \in (-10,9)$. 
Using the fluxes for $\nu_e$, $\bar{\nu}_e$, $\nu_\mu$, and $\bar{\nu}_\mu$ and taking into account that our signal region corresponds to $\cos\theta_z \geq 0.1$, the number of the background events can be obtained in analogy with formula~\eqref{eq:rateDUNE}:
\begin{align}\label{eq:rateNu}
R_{\nu} &= N_{\text{Ar}} \sum_{\nu =\nu_e, \bar{\nu}_e, \nu_\mu, \bar{\nu}_\mu} \int_{E_\nu}  dE_\nu \, \sigma_{\nu\text{Ar}}^{\text{eff}}(E_\nu)\, \int_{0.1}^1 d\cos\theta_z \frac{d\Phi_{\nu}}{dE_\nu d\cos\theta_z} \\
& =  N_{\text{Ar}} \sum_{\nu =\nu_e, \bar{\nu}_e, \nu_\mu, \bar{\nu}_\mu} \int_{E_\nu}  dE_\nu \, \sigma_{\nu\text{Ar}}^{\text{eff}}(E_\nu)\, \sum_{j=1}^9 2\pi\, \frac{d\overline{\Phi}_{\nu}^j}{dE_\nu\,d\Omega} \times 0.1
\end{align}
Again, the effective total neutrino-argon cross section $\sigma_{\nu\text{Ar}}^{\text{eff}}$ takes into account the detection thresholds, see appendix~\ref{secGENIE}.
The analysis predicts 990.4 and 930.2 background neutrino events per year at solar minimum and maximum, respectively. 

Ref.~\cite{Berger:2019ttc} also considers the background due to $\nu_\tau$ charged-current interactions (since the $\tau$ lepton may not be reconstructed if it decays). However, the $\nu_\tau$ background only constitutes 2.8\% of atmospheric neutrinos~\cite{Berger:2019ttc}, making it negligible for our studies. Therefore, this work omits it from consideration as it requires complicated simulations of atmospheric neutrino oscillations for the background calculations. 

The CRDM signal at DUNE represents excess interactions above the neutral current atmospheric neutrino background. Using the same statistical method as in~\cite{Berger:2019ttc}, the significance of the discovery of a certain signal process in a counting experiment is calculated from Ref.~\cite{Cowan:2010js}:
\begin{equation}\label{eqStatSig}
    Z = \sqrt{2 \left[(N_s+ N_b)\log\left(1+\frac{N_s}{N_b}\right) -N_s \right]}\,.
\end{equation}
Here $N_s$ and $N_b$ are the numbers of signal and background events, respectively. This formula assumes large statistics (so that the Poisson distribution can be approximated by the Gaussian one), and it also accounts only for the statistical errors for the number of background events $N_b$. As in~\cite{Berger:2019ttc}, the DUNE sensitivity region after 400 kiloton*years of data is taken at the 2$\sigma$ level. After averaging the solar maximum and minimum results, the number of background events is $N_b=10\, (990.4+930.2)/2 = 9603$. The value $Z=2$ then requires the number of signal events at $N_s=196.7 \equiv \Nthr$.



\section{Generating samples of boosted dark matter interactions in GENIE}\label{secGENIE}

\GENIE is a modular neutrino interaction simulation software intended for neutrino cross section measurements at the GeV-scale~\cite{Berger:2018urf,Berger:2019ttc,Andreopoulos2009rq,Andreopoulos:2015wxa}. In addition to neutrino interaction simulations, \GENIE includes beyond standard model physics, such as boosted DM~\cite{Berger:2018urf}. 
We use the Berger formulation in GENIE to model fermionic DM interactions, specifically the GDM18\_00a\_00\_000 tune with GENIE v3\_04\_00. The standard model configuration sets the sign of all left-handed and right-handed parameters to positive. For the spin-dependent analysis, the left-handed parameters are reversed.

The model simulates both QE scattering and DIS, the latter being a catch-all term that may include interactions not typically called deep inelastic scatters. The flux of boosted DM depends on the differential DM-nucleus cross sections with respect to the energy transfer of the DM component ($\omega_\chi$) and momentum transfer ($\mathrm{Q^{2}}$). These differential cross section are calculated for various kinetic energies of DM and targets. 


To compute differential cross sections, samples with high statistics of interactions must be generated with GENIE, and the differential cross section measured from these samples. The measurement of the differential cross section per bin equates to the number of events in a bin multiplied by the total cross section divided by the number of events and bin size. It is not possible to calculate both differential cross sections of $\omega$ and $\mathrm{Q^{2}}$ analytically within GENIE for the two relevant channels. Therefore, the generation of cross section distributions is computationally expensive, requiring each type of interaction, each kinetic energy of the incident DM particle, and each target to have individualized interaction samples. However, it also provides flexibility for future studies as information on the final state particles, hadronization, and intranuclear interactions are also saved when generating these events.

The sample generation happens for ten targets, each a relevant material in the Earth's crust or a dominant CR element. Targets include hydrogen, helium, carbon, oxygen, magnesium, aluminum-27, silicon, argon-40, potassium-39, and calcium (see appendix~\ref{sec:att} and \sec\ref{sec:CRflux}). Interactions are simulated to calculate the cross sections for DM masses from 50 MeV to 10 GeV, with cross sections for lower masses scaled based on results for 50 MeV. The kinetic energy of the incident DM on the target is set at discrete points and sampled using a logarithmic scale starting at 0.1 GeV and ending at 100 GeV with 40 intermediate steps. Each kinetic energy step for each target and DM mass for each interaction type contains one hundred thousand events. The resulting $\omega$ and $Q^{2}$ from all events in a sample are binned into a differential histogram and saved using ROOT~\cite{BRUN199781, rootRepo}. An interpolating spline is then drawn over each histogram for ease of interpretation. 

The rate of DM events in DUNE~\eqref{eq:rateDUNE} is then estimated from the detector capabilities of DUNE. The DUNE liquid argon detectors are designed for detecting MeV-scale particles produced from neutrino interactions. An event is detected if a pion, proton, or kaon is produced above the kinetic energy threshold for DUNE. The kinetic energy threshold for charged pions is 20 MeV, and the kinetic energy threshold for protons and kaons is 40 MeV. There is no threshold for neutral pions. The thresholds used are from the most recent DUNE results from their ProtoDUNE Single-Phase detector~\cite{DUNE:2024qgl}. The cross section is then re-calculated by multiplying the total boosted DM cross section by the fraction of detectable interactions given those thresholds. The same process is repeated to calculate the atmospheric neutrino background from neutral current interactions~\eqref{eq:rateNu}.

\bibliography{biblio.bib}
\bibliographystyle{JHEP_improved}

\end{document}